\documentclass[draft]{agujournal2019}
\usepackage{amsmath, amssymb, mathtools, subcaption}
\usepackage{algorithm}
\usepackage{algpseudocode}
\usepackage{url} 
\usepackage{lineno}
\usepackage{soul}
\usepackage{bm}
\usepackage[T1]{fontenc}
\usepackage{siunitx}
\DeclareSIUnit\month{mo}
\DeclareSIUnit\millibar{mb}
\DeclareSIUnit\micron{micron}
\DeclareSIUnit\sqrtkelvin{K^{1/2}}
\AtBeginEnvironment{align}{\sisetup{per-mode=fraction}}
\AtBeginEnvironment{align}{\sisetup{per-mode=fraction}}
\AtBeginEnvironment{equation}{\sisetup{per-mode=fraction}}
\graphicspath{{./figures/}}
\draftfalse

\journalname{Journal of Advances in Modeling Earth Systems (JAMES)}

\begin{document}

%
%


\title{Leveraging higher-order time integration methods for improved computational efficiency in a rainshaft model}



%
%




\authors{%
Justin Dong\affil{1},
Sean P. Santos\affil{2},
Steven B. Roberts\affil{1},
Christopher J. Vogl\affil{1},
and Carol S. Woodward\affil{1}
}

\affiliation{1}{Lawrence Livermore National Laboratory, 7000 East Ave, Livermore, CA 94550}
\affiliation{2}{Pacific Northwest National Laboratory, 902 Battelle Blvd, Richland, WA 99354}




\correspondingauthor{Sean P. Santos}{sean.santos@pnnl.gov}



\begin{keypoints}
\item The time integration method commonly used by microphysics schemes requires limiters for stability and produces large approximation errors.
\item Higher-order Runge--Kutta methods with adaptive time steps drastically improve solution accuracy and efficiency compared to current methods.
\item Analysis of microphysical processes in terms of timescales and Jacobian eigenvalues helps explain behavior of time integration methods.
\end{keypoints}

%
%

%
%


\begin{abstract}
Cloud and precipitation microphysics packages in atmospheric general circulation models typically use first-order time integration methods with a large time step, requiring \emph{ad hoc} limiters and substepping of the sedimentation scheme to prevent numerical instability. We investigate alternative methods for rain microphysical processes in the current Energy Exascale Earth System Model (E3SMv3), provided by the Predicted Particle Properties (P3) scheme. Using an offline rainshaft model containing P3's rain microphysics as a proxy for E3SMv3, we find that rain microphysics is underresolved in time at E3SMv3's default \SI{300}{\second} time step. Accurately resolving the rainshaft model processes in time requires an over 100x reduction in time step, increasing wall clock time by 40x.

However, higher-order time integrators based on Runge--Kutta methods offer improved solution accuracy for a given computational cost. Adaptive time stepping is key to obtaining computationally efficient microphysics results, eliminating the need for specialized substepping procedures in the sedimentation process.
An adaptive second-order Runge–Kutta method approximates a high-temporal-resolution reference solution at only \num{2.6}x the cost of the default P3 scheme; this method achieves high accuracy greater than 15x more efficiently than simply reducing the time step of P3 in the rainshaft model.
We also analyze the timescales of the rain processes to obtain insight about the maximum time step each process is able to take while maintaining stability and accuracy, and about how individual processes should be grouped together for most efficient results. 

\end{abstract}

\section*{Plain Language Summary}
Atmospheric models for weather and climate studies simulate cloud microphysics (i.e., the mechanisms by which clouds and precipitation are created and evolve in the atmosphere). Time integration methods enable these models to produce snapshots of the cloud state at a series of points in time. Earth system models produce snapshots infrequently (e.g. every 5 minutes), which is inexpensive but not very accurate. Producing more frequent snapshots is more expensive, requiring more computational resources, but results in more accurate simulations. In this work, we propose replacing current low-accuracy time integration methods for rain microphysics with more sophisticated, high-accuracy methods that achieve improved simulation accuracy, in exchange for a mild increase in compute costs (i.e., small when compared with the full cost of an atmosphere model). These high-accuracy methods are paired with an adaptive procedure for guiding the generation of snapshots; snapshots are produced more frequently when the rain is changing quickly and less frequently when the rain evolves slowly.

\section{Introduction}

Atmospheric general circulation models (AGCMs) typically run at much lower temporal resolution than regional atmospheric models, using time steps on the order of tens of minutes, to ameliorate the high computational cost of running on a global domain for multiple decades or centuries, as is commonly required for climate studies. Given that such long time steps are inadequate to resolve the effects of cloud and precipitation microphysics, most AGCMs ``substep'' their cloud microphysics schemes, with time steps as low as 30-60 seconds often used for models with prognostic precipitation \cite{michibata2019miroc6,sant2015echam5}. Indeed, microphysics time steps of greater than 60 seconds have been repeatedly shown to produce significantly different prognostic precipitation outputs than time steps at or below 60 seconds \cite{posselt2008echam5,chosson2014gem,zheng2020magic,santos2021e3sm}, indicating that time steps above 1 minute are often inadequate to resolve key precipitation processes.

Nonetheless, there are many models that use longer microphysics time steps, such as the ``family'' of AGCMs descended from the Community Atmosphere Model \cite<CAM;>[]{cam5description}, as well as some models that use microphysical parameterizations originally developed for CAM \cite<e.g., GISS ModelE3;>[]{cesana2019modele}. This family includes the Energy Exascale Earth System Model (E3SM) Atmosphere Model \cite{E3SMv1Overview}, which is the focus of this study and uses a \SI{300}{\second} time step for its microphysics. Using long time steps (commonly \SIrange{60}{900}{\second}) would ordinarily result in numerical instability or other pathological behaviors that would crash the model or produce unrealistic outputs. These outcomes are avoided by (a) substepping sedimentation of hydrometeors at a smaller, typically adaptive, time step than the overall microphysics time step, and (b) the use of limiters to ensure nonnegativity of hydrometeor concentrations and a reasonable range of particle sizes. This approach results in a computationally cheap and robust scheme at the cost of reduced fidelity: temporal resolution is coarse, and the evolution of the systems may often be controlled by limiter ``guardrails'' rather than physical processes. Any biases in the resulting model may be compensated by retuning the parameterizations, but since these changes do not address the numerical error underlying the biases, the retuning may need to be repeated whenever the time step is changed.

Even for models willing to accept the computational cost of running microphysics at a 60 second time step, there may be regimes where time steps as short as 10 seconds are required to well-resolve some processes at typical AGCM spatial resolutions \cite{santos2020mg2}, and even shorter time steps may be required at convection-permitting resolutions \cite{barrett2019timestep}. Resolving such processes might seem to require an order of magnitude increase in computational cost, insofar as reducing the model time step would increase the cost proportionally. However, most parameterizations of cloud and precipitation microphysics use simple first-order, explicit time integration methods, particularly forward Euler for most individual microphysical processes and sequential operator splitting for process coupling. These methods are easy to implement, but they are not necessarily the most computationally efficient options, so we should consider whether there are more efficient methods for accurately calculating the effects of microphysical processes. For many applications, first-order methods are less efficient than higher-order methods as measured by wall clock time to reach a desired solution accuracy, particularly when leveraging adaptive time steps.

The accuracy of time integration is less important when other sources of error are assumed to dominate the model uncertainty (e.g. structural or parametric error in the physics formulation, coarse spatial resolution), and implementing new time integration methods requires extra development time. However, time discretization error is not always small compared to other sources of uncertainty, even for models with coarse spatial resolution and highly uncertain parameterizations (e.g. of convection). Furthermore, while global cloud-resolving models show great promise in reducing uncertainties related to convection and spatial resolution, this reduction comes with a very high computational cost, so it is tempting to retain relatively coarse time steps in the physics. For instance, the Simple Cloud-Resolving E3SM Atmosphere Model uses a horizontal resolution of $\sim$\SI{3.25}{\kilo\meter}, over 30x finer than the ``low'' resolution ($\sim$\SI{110}{\kilo\meter}) dynamics grid used in E3SMv3 and over 8x finer than that of the ``high'' resolution ($\sim$\SI{27}{\kilo\meter}) version, but it uses a \SI{100}{\second} time step for the microphysics, only a 3-fold increase in temporal resolution over the lower-resolution configurations \cite{donahue2024scream}. If AGCMs continue to allocate computational resources mainly toward higher spatial resolution rather than higher temporal resolution, we should expect inaccurate time integration to make up a larger share of the numerical errors affecting model uncertainty.

To avoid the aforementioned extra development time to implement second- and higher-order methods, we leverage SUNDIALS, the SUite of Nonlinear and DIfferential/ALgebraic equation Solvers \cite{hindmarsh2005sundials}, to provide a wide range of alternative, high-order time integration schemes.
SUNDIALS is a C library containing six packages; however, our use is limited to the ARKODE \cite{reynolds2023arkode} package which contains implementations of one-step time integration methods. In order to apply these methods more easily for microphysics problems, we have created a new software framework called the Sedimentation and microPhysics Accurate and Efficient Coupler/Integrator for Earth Systems (SPAECIES). SPAECIES allows the user to implement a new microphysical process by writing a method that accepts the atmospheric state of a column (possibly including additional diagnostic variables) and outputs the rate of change of prognostic variables in that column. SPAECIES then allows multiple process objects to be combined and integrated in various ways using the methods provided by SUNDIALS. When using SUNDIALS methods, microphysical processes are handled via a method-of-lines approach that decouples the time and space discretizations, allowing us to easily swap in different time integrators and run extensive performance comparisons, though SPAECIES is flexible enough to allow other approaches (such as those currently used by most microphysics parameterizations).

Our work consists of two major contributions. The first is that we examine the efficiency of several higher-order time integration methods applied to a rainshaft model based on a subset of microphysical processes in the Predicted Particle Properties (P3) scheme \cite{morrison2015parameterization}. We demonstrate that the current time integration method in P3, while stable at E3SMv3's default time step (\SI{300}{\second}) due to the use of positivity and size limiters (defined in Section \ref{sec:P3}), produces highly inaccurate solutions of the rainshaft model at such coarse temporal resolutions. Running on E3SMv3 inputs using the default vertical and temporal resolution from that model, we compare the solution from the current method taking a single \SI{300}{\second} time step to a reference solution using high-temporal-resolution substepping.
There is a \num{78}\% relative difference in rain mass mixing ratio between the two, measured by weighted root mean square error.
This difference does not meaningfully decrease until the model time step is reduced by more than a factor of \num{100}. Furthermore, considering the reduction in time step necessary to adequately resolve important solution features like sharp gradients, the corresponding wall clock time to solution would increase by a factor of \num{40}. We ask whether we can achieve better accuracy than the P3 time integration method without drastically increasing the wall clock time to solution. To this end, we consider several explicit Runge--Kutta (ERK) methods, diagonally implicit Runge--Kutta (DIRK) methods, implicit-explicit (ImEx) Runge--Kutta methods, higher-order operator splitting methods, and multirate infinitesimal generalized additive Runge--Kutta (MRI-GARK) methods. All ERK, DIRK, ImEx, and MRI-GARK methods considered support adaptive time stepping.

The second major contribution of this work is to provide a rigorous analysis of several microphysical processes in order to determine the impact of partitioning, if any, on time integrator performance. The P3 scheme uses sequential splitting to evaluate processes that do not couple spatially-separated quantities (``local'' processes) separately from sedimentation, which simplifies the application of limiters to guarantee positivity on the one hand, while also simplifying substepping of the sedimentation to satisfy the Courant--Friedrichs--Lewy (CFL) condition on the other. Several of the other time integration methods we consider also rely on an appropriate partitioning of the microphysical processes based on mathematical properties such as numerical stiffness and/or physical properties such as the timescale over which the process occurs.

Although the results of our analysis are specific to a particular modeling context (warm rain handled by P3 within E3SM), our approach can be applied to other processes beyond those in the P3 rain microphysics. Namely, by separating out the implementation of physical processes from any time integration or limiter implementation, it becomes possible to explore a much wider range of time integration methods. Furthermore, thoughtful consideration of characteristic time scales of different processes can provide insight into why certain methods may be more accurate or efficient than others. 

The remainder of this work is structured as follows. In Section \ref{sec:rainshaft model}, we describe the minimal rainshaft model we use to test time integration methods, as well as the time integration method currently used in P3. A stability and timescale analysis of microphysical processes is presented in Section \ref{sec:process analysis}. In Section \ref{sec:time integration}, we describe the various higher-order time integration methods we have implemented for P3's rain microphysics, and we provide results demonstrating their efficiency in the rainshaft model. Conclusions follow in Section \ref{sec:conclusion}.



%
%

%
%
\section{Rainshaft Model} \label{sec:rainshaft model}

For this study, we focus on a subset of the P3 microphysics \cite{morrison2015parameterization} that is mathematically simple enough that analysis of microphysical processes and time integration methods under consideration is practical, without oversimplifying the physics to the point that our model becomes a poor proxy for the P3 physics as a whole. To accomplish this, we focused on the meteorological regime of warm rain beneath cloud base. There are three P3 processes that act solely on rain, namely a rain coalescence and breakup process (which we will call ``self-collection'' for short), rain evaporation, and rain sedimentation. Thus, a model that contains only these three processes is equivalent to the full P3 model so long as cloud-sized drops and ice are not present and not actively being formed. Such ``rainshaft'' models are thus commonly used as proxies for full bulk microphysics models like P3, e.g. in \citeA{paukert2019} and \citeA{morrison2020rainshaft}.

We generate warm rain scenarios by filtering data from E3SMv3 runs to select parts of the domain where only rain is present and use this data to form the initial and boundary conditions of our simulations. In this regime, the main differences in physics between our rainshaft model and the P3 model as actually implemented in E3SMv3 are:
\begin{enumerate}
    \item Our model has a more limited domain (only reaching up to the cloud base, and not simulating the cloud itself), so we assume a fixed upper-boundary condition for rain mass/number at cloud base in our model, rather than prognosing these values.
    \item Since our model does not contain ice, we do not implement freezing processes for supercooled rain. However, our dataset contains no grid cells where homogeneous freezing would occur, only 0.9\% of grid cells have temperatures low enough for P3's immersion processes to occur (below \SI{-4}{\celsius}), and only 0.3\% of columns maintain these low temperatures all the way to the surface.
\end{enumerate}

The rainshaft model equations implementing the P3 rain microphysics are given by
\begin{equation} \label{eq:rainshaft model}
    \begin{dcases}
        \frac{\partial T}{\partial t} = -\frac{L_{\text{v}}}{c_{p}} S_{q,\text{evap}}(T,q,n_{\text{r}},q_{\text{r}}) \\
        \frac{\partial q}{\partial t} = S_{q,\text{evap}}(T,q,n_{\text{r}},q_{\text{r}}) \\
        \frac{\partial n_{\text{r}}}{\partial t} = \frac{1}{\rho(T,q)}\frac{\partial \left(v_{n_{\text{r}}}(T,q,n_{\text{r}},q_{\text{r}})n_{\text{r}}\rho(T,q)\right)}{\partial z} - S_{n,\text{evap}}(T,q,n_{\text{r}},q_{\text{r}}) - S_{\text{rsc}}(T,q,n_{\text{r}},q_{\text{r}})\\
        \frac{\partial q_{\text{r}}}{\partial t} = \frac{1}{\rho(T,q)}\frac{\partial \left(v_{q_{\text{r}}}(T,q,n_{\text{r}},q_{\text{r}})q_{\text{r}}\rho(T,q)\right)}{\partial z} - S_{q,\text{evap}}(T,q,n_{\text{r}},q_{\text{r}}).
    \end{dcases}
\end{equation}
The prognostic variables are the temperature $T$, water vapor mixing ratio $q$, raindrop number concentration $n_{\text{r}}$, and rain mass mixing ratio $q_{\text{r}}$. The derived variables $\rho$, $v_{n_{\text{r}}}$, and $v_{q_{\text{r}}}$ are the local density of air and fall speeds for rain number and mass, respectively. P3's physics is formulated in terms of ``dry'' mixing ratios ($n_{\text{r}}$ and $q_{\text{r}}$ are ratios over dry mass), so we define the density $\rho$ to also include only dry mass:
\begin{align}
    \rho(T,q) = \frac{p}{R_{\text{d}} T (1 + q/\epsilon_{\text{H2O}})},
\end{align}
where the constants $R_{\text{d}}$, and $\epsilon_{\text{H2O}}$ are, respectively, the gas constant for dry air and the molecular mass ratio of water vapor over dry air. Note that $p$ is the total (\emph{not} dry partial) pressure. Pressure is constant over time for each grid cell since the grid cells are defined by our adaptation of the E3SM grid. The specific microphysical processes applied to the rainshaft model in this work are sedimentation, raindrop self-collection denoted by $S_{\text{rsc}}$, and liquid evaporation denoted by $S_{n,\text{evap}}$ and $S_{q,\text{evap}}$. Full details of these processes may be found in \ref{app:processes}, with the values for all physical constants found in Table \ref{tab:constants}.

%
%
\subsection{Spatial discretization} \label{sec:spatial}

Following the spatial discretization of the P3 scheme \cite{morrison2015parameterization}, the sedimentation terms in Equation \ref{eq:rainshaft model} are discretized in space using a first-order (upwind) finite difference formulation that guarantees rain mass and number are locally conserved (after accounting for fluxes across the model boundaries). This discretization results in the semidiscrete system of equations
\begin{align} \label{eq:semidiscrete}
    \begin{dcases}
        \frac{d T_{i}}{d t} = -\frac{L_{\text{v}}}{c_{p}} S_{q,\text{evap}}(T_{i},q_{i},n_{\text{r},i},q_{\text{r},i}) \\
        \frac{d q_{i}}{d t} = S_{q,\text{evap}}(T_{i},q_{i},n_{\text{r},i},q_{\text{r},i}) \\
        \frac{d n_{\text{r},i}}{d t} = \frac{1}{\rho_i}\left( \frac{v_{n_{\text{r}}}(\rho_{i-1},n_{\text{r},i-1},q_{\text{r},i-1})n_{\text{r},i-1}\rho_{i-1} - v_{n_{\text{r}}}(\rho_{i},n_{\text{r},i},q_{\text{r},i})n_{\text{r},i}\rho_{i}}{\Delta z_{i}} \right)\\
        \;\;\;\;\;\;\;\;\;\;\;\;\;\;\;\;\;\;\;\;\;\;\;\;\;\;- S_{n,\text{evap}}(T_{i},q_{i},n_{\text{r},i},q_{\text{r},i}) - S_{\text{rsc}}(T_{i},q_{i},n_{\text{r},i},q_{\text{r},i})\\
        \frac{d q_{\text{r},i}}{d t} = \frac{1}{\rho_{i}}\left( \frac{v_{q_{\text{r}}}(\rho_{i-1},n_{\text{r},i-1},q_{\text{r},i-1})q_{\text{r},i-1}\rho_{i-1} - v_{q_{\text{r}}}(\rho_{i},n_{\text{r},i},q_{\text{r},i})q_{\text{r},i}\rho_{i}}{\Delta z_{i}} \right)\\
        \;\;\;\;\;\;\;\;\;\;\;\;\;\;\;\;\;\;\;\;\;\;\;\;\;\;- S_{q,\text{evap}}(T_{i},q_{i},n_{\text{r},i},q_{\text{r},i}),
    \end{dcases}
\end{align}
for $i = 1, \dots, N_{z}$ where $N_{z}$ is the number of cells in the spatial grid. Consistent with E3SM conventions, $z$ values are altitudes measured from the ground. However, indexing begins with the highest altitude grid cell so that $z_0$ is the altitude of the model top and $z_{N_{z}}=0$ is altitude of the interface at the Earth's surface. For convenience, the symbol $\rho_i$ is used to denote $\rho(T_i,q_i)$. We denote by $\Delta z_{i}$ the length of the cell $(z_{i-1},z_{i})$ (i.e., $\Delta z_{i} = z_{i-1}-z_{i}$). At the upper boundary, we must calculate incoming fluxes of $n_{\text{r}}$ and $q_{\text{r}}$, which is done by specifying $\rho_0$, $n_{\text{r},0}$ and $q_{\text{r},0}$ as boundary conditions for each column. These values, as well as the number of grid cells in the domain $N_{z}$, are determined based on the location and properties of the cloud base from E3SMv3 data, as described in the next section. 

For ease of notation, we occasionally refer to the semidiscrete system Equation \ref{eq:semidiscrete} in the more compact form
\begin{align} \label{eq:semidiscrete operator form}
    \frac{d\bm{y}(t)}{d t} = \bm{\mathfrak{F}}(\bm{y}(t)), \quad\bm{y}(t) = (\bm{T}(t), \bm{q}(t), \bm{n}_{\text{r}}(t), \bm{q}_{\text{r}}(t))^{T},
\end{align}
where $\bm{X}(t) = (X_{1}(t), \dots, X_{N_z}(t))^{T}$ for $X = T, q, n_{\text{r}}, q_{\text{r}}$, and $\bm{\mathfrak{F}}$ is the rate of change of the spatially-discretized prognostic variables.

%
%
\subsection{Boundary and initial conditions} \label{sec:boundary conditions}

Since boundary and initial condition data is taken from the E3SM Atmosphere Model version 3 (EAMv3), we will briefly discuss time integration in that model. At the standard ($\sim\SI{1}{\degree}$) resolution, the model uses an overall time step of \SI{1800}{\second}, which is the time step used to couple the atmospheric physics, atmospheric dynamics, and non-atmosphere model components. Different atmospheric processes are generally coupled using sequential splitting, e.g. dynamics is evaluated before deep convection, then other cloud physics (including microphysics), then radiation. The parameterization that handles macrophysics, shallow convection, and unresolved turbulence (especially in the boundary layer) is the Cloud Layers Unified By Binormals scheme (CLUBB). CLUBB, droplet activation (performed by the modal aerosol parameterization outside of P3), and P3 are substepped in the same loop together, with a \SI{300}{\second} sub-time step. That is, within each atmospheric time step, six smaller time steps are taken that consist solely of CLUBB, the droplet activation scheme, and P3 being run sequentially, in that order.

Boundary and initial conditions for our rainshaft model experiments are derived from a \num{15}-month global run of EAMv3, coupled to a data ocean under present-day conditions. A global snapshot of all inputs to P3's rain microphysics is captured every \num{30} days and \num{10} hours (i.e., every 365/12 days). Although this is not a perfectly uniform random sampling of P3 inputs, using a non-integer number of days does guarantee that for any given point on the planet, multiple points in the diurnal cycle are sampled during each season. The snapshots also save inputs for each of the six CLUBB/P3 substeps that occur during the selected time step. We include data from multiple CLUBB/P3 substeps because it is expected that the first substeps within every EAMv3 time step will systematically differ from later substeps. The first three snapshots (roughly two months) are treated as model spinup and discarded, and the remaining \num{12} snapshots thus contain a sampling of \num{1555200} columns of P3 inputs taken roughly uniformly across locations, seasons, times of day, and P3 substeps.

The vast majority of these columns did not correspond to the meteorological conditions which the rainshaft model is intended to address (i.e., no rain is present or there is significant ice). To focus on liquid precipitation produced from boundary-layer clouds, we select atmospheric columns from the dataset with the following criteria:
\begin{itemize}
    \item A cloud base exists between 400 and 2000 meters, where cloud base is defined as the lowest-altitude interface between grid cells such that cloud liquid is present in the grid cell above and not in the grid cell below.
    \item Rain is present in the lowest cloud level, signifying recent or ongoing rain production.
    \item Ice is not present in the lowest cloud level or below cloud base.
\end{itemize}
The presence or absence of cloud liquid, rain, or ice was determined from mass mixing ratios (grid-cell mean, \emph{not} in-cloud values), requiring a value above \SI{1.e-6}{\kilo\gram\per\kilo\gram} for the corresponding hydrometeor to be considered ``present''. This value is chosen partly for consistency with internal limiters in E3SM and partly because \SI{1.e-6}{\kilo\gram\per\kilo\gram} corresponds to a local minimum in the distribution of the logarithm of hydrometeor masses (noted in Figures S1 and S2 from supplemental information).

From the columns matching these criteria, we randomly select $N_{\text{col}} = 1000$ columns for this study, with the E3SM grid cells below cloud base supplying the initial conditions for our experiments. E3SM uses a maximum-overlap assumption where the precipitation fraction is considered constant below cloud base (except that it is set to the minimum allowed value of \num{1.e-4} in grid cells where rain and ice mass are both below \SI{1.e-14}{\kilo\gram\per\kilo\gram} and in all lower cells). The initial conditions for $q_{\text{r}}$ and $n_{\text{r}}$ are determined by dividing the grid-cell-mean values from EAMv3 by rain area fraction to obtain below-cloud mean values. The rainshaft model then works entirely with quantities inside the precipitating area, removing the need to convert back to grid-cell-mean values during simulation. Initial values of $q$ and $T$ are taken directly from the grid-cell-mean values in EAM. Upper boundary conditions for $q_{\text{r}}$ and $n_{\text{r}}$ are chosen to match in-cloud values in the grid cell directly above the cloud base, providing a constant source of rain via sedimentation. Upper boundary conditions for $q$ and $T$ are also taken from the same grid cell, but are only used to calculate air density $\rho$ at the upper interface (since density affects the fall speed, and hence flux, at the model top).

\subsection{The original P3 time integration method} \label{sec:P3}

The original P3 code \cite{morrison2015parameterization,morrison2015parameterization2} and the modified version used in E3SM both rely on first-order explicit methods. Specifically, sequential splitting, also referred to as Lie--Trotter splitting, is used to apply the effects of all local microphysical processes first, followed by sedimentation. These local processes are not sequentially split from one another, but rather applied together with a single forward Euler step.

Processes involving phase changes to/from water vapor (including rain evaporation) are an exception. Those processes use what is called the ``semianalytic'' or ``quasianalytic'' method, in which a linear PDE that approximates the evolution of supersaturation over short periods is obtained, this linear PDE is solved analytically to obtain supersaturation as a function of time, and process rates are derived from this supersaturation \cite{clark1973,morrison2008,morrison2015parameterization}. This semianalytic method is an example of what is called an ``exponential integrator'' in numerical analysis, in this case developed to improve the accuracy and stability of the model when using time steps above \SI{1}{\second} \cite{morrison2008}. In the context of the rainshaft model where only the evaporation of rain affects relative humidity, this exponential integrator approach is equivalent to applying the forward Euler method with the rate of change multiplied by the weighting factor
\begin{align} \label{eq:evap weight}
    w_{\Delta t} &= \frac{1 - e^{-\Delta t/\tau}}{\Delta t/\tau},
\end{align}
with $\Delta t$ being the microphysics time step and $\tau$ being a timescale associated with water vapor diffusion (see Table \ref{tab:evap equations}).

Both the forward Euler method and the exponential integrator can produce numerical instabilities and unphysical states due to the long model time step (e.g. evaporating so much rain water in one step that $q_{\text{r}}$ becomes negative), but this is prevented in P3 by a set of limiters that scale down process rates that would otherwise produce negative values, ensuring that all mass mixing ratios and number concentrations remain nonnegative. We refer to these adjustments as the ``positivity limiters'', noting that significant care is taken in P3 to implement these in such a way as to ensure conservation of mass. The value of $n_{\text{r}}$ is also clipped by further limiters designed to ensure that the mean drop size is kept in the range \SIrange{10}{500}{\micro\meter}, by changing $n_{\text{r}}$ without changing $q_{\text{r}}$. We refer to these limiters as the ``size limiters''. Technical descriptions of the positivity and size limiters are provided in \ref{app:limiters}.

As described in Section \ref{sec:spatial}, sedimentation is discretized in space with a first-order upwind approximation. Preventing negative values (and related pathologies) resulting from evolving sedimentation with forward Euler requires substepping the process finely enough to satisfy the CFL condition, which is approximated in P3 by $\Delta t < \Delta z_i / v_{q_{\text{r}}}(n_{\text{r},i},q_{\text{r},i})$ for $i \in \{1,\dots\, N_z\}$. (Further discussion of the CFL condition may be found in Section \ref{sec:sedimentation}.) Because this condition depends on $v_{q_{\text{r}}}$, which depends on the hydrometeor variables themselves, it needs to be reevaluated, and the time step potentially adjusted, after every substep. A summary of the P3 time integration method may be found in Algorithm \ref{alg:P3}.
\begin{algorithm}
\caption{The P3 time integration method applied to the rainshaft model.}\label{alg:P3}
\begin{algorithmic}[1]
\Require Initial condition $\bm{y}_{0}$, time step $\Delta t$ ($=\SI{300}{\second}$ in E3SM), initial time $t_{0}$, and final time $t_{\text{f}}$.
\State Set $n = 0$
\While{$t_n < t_{\text{f}}$}
    \State $\Delta \widetilde{t} = \min\{ \Delta t, t_{\text{f}}-t_{n} \}$
    \State Compute weight $w_{\Delta \widetilde{t}}$ from Equation \ref{eq:evap weight} using $\bm{y}_{n}$.
    \State Apply evaporation and self-collection:
    \begin{equation*}
        \bm{y}^{(1)} = \bm{y}_{n} + \Delta \widetilde{t} \cdot \Big(w_{\Delta \widetilde{t}}\mathfrak{F}_{\text{evap}}(\bm{y}_{n}) + \mathfrak{F}_{\text{rsc}}(\bm{y}_{n})\Big).
    \end{equation*}
    \State Apply positivity and size limiters: $\bm{y}^{(1)} \gets \mathfrak{L}_{\text{pos}}(\bm{y}^{(1)})$ and $\bm{y}^{(1)} \gets \mathfrak{L}_{\text{size}}(\bm{y}^{(1)})$
    \State Set $t_{\text{sub}} = t_{n}$ and $\bm{y}_{\text{sub}} = \bm{y}^{(1)}$.
    \While{$t_{\text{sub}} < t_{n} + \Delta \widetilde{t}$} \Comment{Sedimentation substepping.}
        \State Compute time step $(\Delta t)_{\text{sed}}$ based on Equation \ref{eq:P3 cfl}.
        \State Set $(\Delta t)_{\text{sub}} = \min\{ (\Delta t)_{\text{sed}}, t_{n} + \Delta \widetilde{t}-t_{\text{sub}} \}$.
        \State Apply sedimentation with substep and limiter:
        \begin{align*}
            \bm{y}_{\text{sub}} \gets \bm{y}_{\text{sub}} + (\Delta t)_{\text{sub}} \cdot \mathfrak{F}_{\text{sed}} (\bm{y}_{\text{sub}}).
        \end{align*}
        \State Apply positivity and size limiters: $\bm{y}_{\text{sub}} \gets \mathfrak{L}_{\text{sedpos}}(\bm{y}_{\text{sub}})$ and $\bm{y}_{\text{sub}} \gets \mathfrak{L}_{\text{size}}(\bm{y}_{\text{sub}})$.
        \State $t_{\text{sub}} \gets t_{\text{sub}} + (\Delta t)_{\text{sub}}$.
    \EndWhile
    \State Set $\bm{y}_{n+1} = \bm{y}_{\text{sub}}$, $t_{n+1} \gets t_{n} + \Delta \widetilde{t}$, and $n \gets n+1$.
\EndWhile\\
\Return $\bm{y}_{n}$
\end{algorithmic}
\end{algorithm}

\subsubsection{Accuracy of the P3 time integration method} \label{sec:P3 accuracy}

\begin{figure}[ht!]
    \centering
    \includegraphics[width=0.75\linewidth]{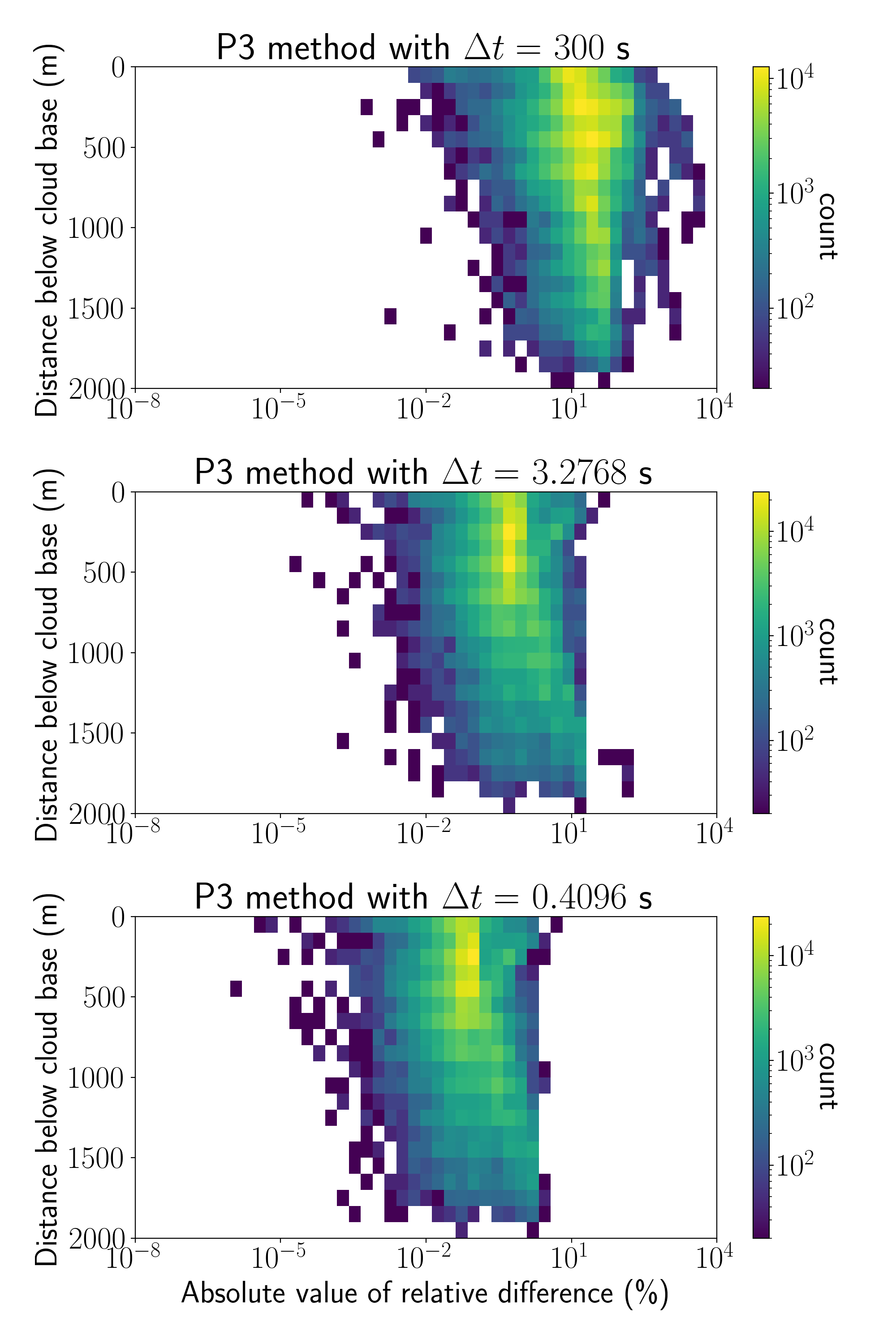}
    \caption{2D histogram of pointwise absolute value relative difference in $q_\text{r}$ between highly-resolved reference and P3 solution with time steps of \SI{300}{\second}, \SI{3.2768}{\second}, and \SI{0.4096}{\second}, with the differences shown as percentages relative to the highly-resolved reference solution. The colormap shows the number of grid points per $xy$ bin. The exact error metric used in the binning is $|(q_{\text{r},i} - q_{\text{r,ref},i})/q_{\text{r,ref},i} \times 100|$, where $q_\text{r,ref}$ is the highly-resolved reference.}
    \label{fig:P3 histogram}
\end{figure}

\begin{figure}[ht!]
    \centering
    \includegraphics[width=0.99\linewidth]{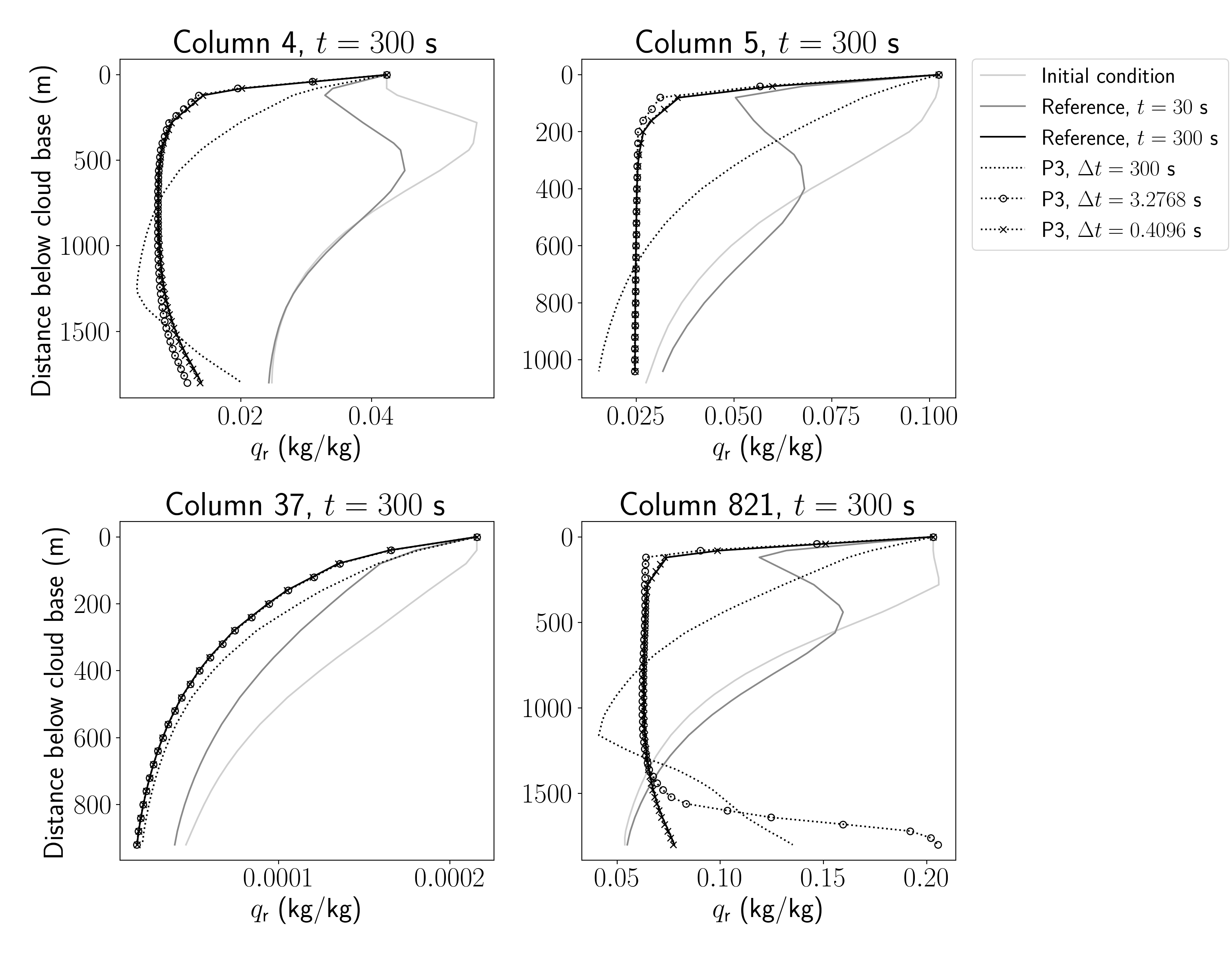}
    \caption{Selected solution profiles at \SI{300}{\second} using the P3 time integration method with $\Delta t = \SI{300}{\second}$, \SI{3.2768}{\second}, and \SI{0.4096}{\second}. The initial condition from E3SM data is also shown for comparison as well as the highly-resolved reference solution at $t=\SI{30}{\second}$ and $\SI{300}{\second}$.}
    \label{fig:P3 profiles}
\end{figure}

To motivate the need for more efficient time integration methods, we first examine the accuracy of the P3 time integration method at its default \SI{300}{\second} time step applied to the rainshaft model (Equation \ref{eq:rainshaft model}). Figure \ref{fig:P3 histogram} shows the distribution of pointwise relative differences expressed as percentages between the solution obtained using the P3 method compared with a highly-resolved reference solution (details of the reference solution may be found in Section \ref{sec:time integration}). Selected $q_{\text{r}}$ solution profiles at \SI{300}{\second} are provided in Figure \ref{fig:P3 profiles}. Results using the P3 method with significantly smaller time steps of \SI{3.2768}{\second} and \SI{0.4096}{\second} are also provided for comparison.

We observe that a large number of pointwise relative differences exceed $10\%$ and even $100\%$ when the default time step of \SI{300}{\second} is taken. Moreover, many of the solution profiles are highly inaccurate and fail to capture vital solution behavior such as sharp gradients near the cloud base in columns 5 and 821. When the time step is reduced to \SI{3.2768}{\second}, the typical pointwise relative differences are reduced to $0.1\%$ -- $10\%$, but there are many outliers, such as column 821 at the Earth's surface, for which the P3 method produces a physically unrealistic peak where none exists in the highly-resolved reference solution. If one further reduces the time step to \SI{0.4096}{\second}, then the largest pointwise relative differences are around $1\%$ and the solution profiles capture the correct physical behavior. However, this significant reduction in time step increases the simulation wall clock time by nearly a factor of \num{40}, a highly undesirable outcome considering the current computational costs of running a typical AGCM. Thus, when evaluating alternative time integration methods in Section \ref{sec:time integration}, our goal is to better resolve important solution features without drastically increasing P3's computational cost.

\subsection{Regularization of discontinuities} \label{sec:regularization}

The nonlinear nature of the rainshaft model results in a nonlinear system to be solved when any processes are to be integrated in time implicitly. Discontinuities can delay or prevent nonlinear solvers from finding a solution, especially if the nonlinear solver leverages Jacobian information (e.g., Newton's method), and can also degrade convergence rates of the time integration methods. To address these issues, we regularize the mean diameter and evaporation tendency and use the regularized expressions in all numerical experiments in this work.

The inverse mean diameter $\lambda_{\text{r}}$, utilized as an intermediary variable in all processes, contains a jump discontinuity at $q_{\text{r}} = q_{\text{small}}$ due to the limiter imposed when $q_{\text{r}} \leqslant q_{\text{small}}$. Adding a small tolerance to the denominator of the non-zero branch and removing the limiter entirely resolve this continuity issue. Full details are provided in Appendix \ref{app:lambdar regularization}.

In the case of evaporation, the tendency is continuous in its function values but contains a jump discontinuity in its derivative with respect to $q$ due again to the use of limiters. A polynomial interpolant with the desired continuity properties is used to replace the original tendency in a small region around the original point of discontinuity. Full details are provided in Appendix \ref{app:evap regularization} along with numerical experiments demonstrating that the proposed regularizations have negligible effect on solution behavior.

%
%
\section{Process Analysis}
\label{sec:process analysis}

Many of the time integration methods we utilize in Section \ref{sec:time integration} allow different microphysical processes to be treated with different methods and/or time steps. These methods provide the most benefit when the user has some \emph{a priori} knowledge of (i) the timescales associated with each process and (ii) how those timescales, along with the eigenvalues of the discretized processes, affect the stability of the integration methods.

%
%
\subsection{Dimensional analysis and timescales} \label{sec:timescales}

Let $\tilde{t}$, $\tilde{z}$, $\tilde{T}$, $\tilde{q}$, $\tilde{n}_{\text{r}}$, $\tilde{q}_{\text{r}}$, $\tilde{S}_{n,\text{evap}}$, $\tilde{S}_{q,\text{evap}}$, $\tilde{v}$, $\tilde{S}_\text{rsc}$, and $\tilde{\rho}$ denote a characteristic time scale, length scale, temperature, water vapor mixing ratio, raindrop number concentration, rain mass mixing ratio, rain evaporation rate, sedimentation speed, raindrop self-collection rate, and air density, respectively.
Introduce $x^* = x / \tilde{x}$, where $x$ corresponds to one of $t$, $z$, $T$, $q$, $n_{\text{r}}$, $q_{\text{r}}$, $S_{n,\text{evap}}$, $S_{q,\text{evap}}$, $S_\text{rsc}$, or $\rho$.
Similarly, let $v_x^* = v_x / \tilde{v}$, where $x$ corresponds to one of $n_{\text{r}}$ or $q_{\text{r}}$.
Expressing the rainshaft model (Equation \ref{eq:rainshaft model}) in the $x^*$ and $v_x^*$ quantities results in a non-dimensional form of Equation \ref{eq:rainshaft model}:
\begin{equation} \label{eq:nondimensional}
    \begin{dcases}
        \frac{\partial T^*}{\partial t^*} = -\frac{ \frac{L_{\text{v}}}{c_{p}} \tilde{S}_{q,\text{evap}} / \tilde{T}}{ 1/\tilde{t}} S_{q,\text{evap}}^* \\
        \frac{\partial q^*}{\partial t^*} = \frac{ \tilde{S}_{q,\text{evap}} / \tilde{q}}{1/\tilde{t}} S_{q,\text{evap}}^* \\
        \frac{\partial n_{\text{r}}^*}{\partial t^*} = \frac{\tilde{v}/\tilde{z}}{1/\tilde{t}} \frac{1}{\rho^*}\frac{\partial \left(v_{n_{\text{r}}}^* n_{\text{r}}^* \rho^* \right)}{\partial z^*} - \frac{\tilde{S}_{n,\text{evap}} / \tilde{n}_{\text{r}}}{1/\tilde{t}} S_{n,\text{evap}}^* - \frac{\tilde{S}_\text{rsc} / \tilde{n}_{\text{r}}}{1/\tilde{t}} S_\text{rsc}^* \\
        \frac{\partial q_{\text{r}}^*}{\partial t^*} = \frac{\tilde{v} / \tilde{z}}{1/\tilde{t}} \frac{1}{\rho^*} \frac{\partial \left(v_{q_{\text{r}}}^* q_{\text{r}}^* \rho^* \right)}{\partial z^*} - \frac{\tilde{S}_{q,\text{evap}}/\tilde{q}_{\text{r}}}{1/\tilde{t}} S_{q,\text{evap}}^*.
    \end{dcases}
\end{equation}
The various timescales in the system are now obtained by considering how $\tilde{t}$ appears in the non-dimensional coefficients.
The particular timescales of interest are
\begin{itemize}
  \item $\tilde{t} = \frac{\tilde{q}_{\text{r}}}{\tilde{S}_{q,\text{evap}}}$: the timescale on which liquid rain evaporation significantly affects the rain mass mixing ratio (note this timescale is equivalent to $\tilde{t} = \frac{\tilde{n}_{\text{r}}}{\tilde{S}_{n,\text{evap}}}$ because of the proportionality assumption in P3, i.e., Equation \ref{eq:evaporation-number-rate})

  \item $\tilde{t} = \frac{\tilde{n}_{\text{r}}}{\tilde{S}_\text{rsc}}$: the timescale on which raindrop self-collection significantly affects the raindrop number concentration.

  \item $\tilde{t} = \frac{\tilde{z}}{\tilde{v}}$: the timescale on which sedimentation affects the system on the characteristic length scale.
\end{itemize}
Note that the characteristic length $\tilde{z}$ and sedimentation speed $\tilde{v}$ have yet to be specified.
\begin{figure}[t!]
    \centering
    \includegraphics[width=0.99\linewidth]{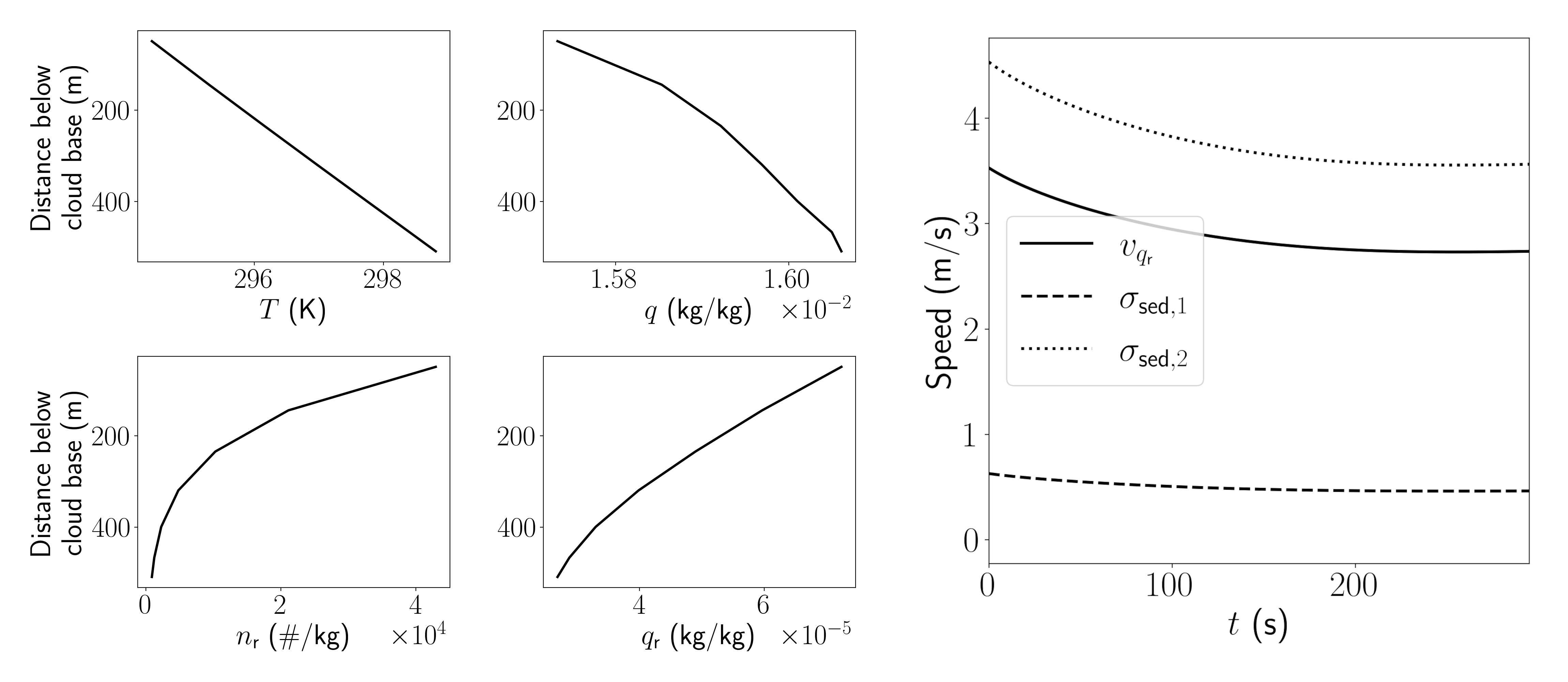}
    \caption{Left: initial solution profiles for temperature, water vapor mixing ratio, raindrop number concentration, and rain mass mixing ratio obtained from a single column of input to P3. Right: corresponding fall speed $v_{q_{\text{r}}}$ and characteristic speeds $\sigma_{\text{sed},1}$ and $\sigma_{\text{sed},2}$. The fall speed used in P3's substepping calculations $(v_{q_{\text{r}}})$ is notably slower than the speed dictated by the largest sedimentation eigenvalue.}
    \label{fig:speeds}
\end{figure}

Traditionally, $v_{q_{\text{r}}}$ is considered the characteristic sedimentation speed $\tilde{v}$, because it is the faster of the two fall speeds $v_{n_{\text{r}}}$ and $v_{q_{\text{r}}}$.
However, the rainshaft model (Equation \ref{eq:rainshaft model}) can be derived from the following system of hyperbolic conservation laws:
\begin{equation} \label{eq:conservative form}
    \begin{dcases}
        \frac{\partial}{\partial t}(\rho T) = - \rho \frac{L_{v}}{c_p} S_{q,\text{evap}} \\
        \frac{\partial}{\partial t}(\rho q) = \rho S_{q,\text{evap}} \\
        \frac{\partial}{\partial t}(\rho n_{\text{r}}) - \frac{\partial}{\partial z} \left(v_{n_{\text{r}}} \rho n_{\text{r}} \right) = -\rho S_{n,\text{evap}} -\rho S_\text{rsc}  \\
        \frac{\partial}{\partial t}(\rho q_{\text{r}}) - \frac{\partial}{\partial z} \left(v_{q_{\text{r}}} \rho q_{\text{r}} \right) = - \rho S_{q,\text{evap}}.
    \end{dcases}
\end{equation}
The speed at which the overall solution propagates in such a hyperbolic system is the maximum hyperbolic characteristic speed \cite{leveque2002finite}, not the fall speeds $v_{n_{\text{r}}}$ and $v_{q_{\text{r}}}$.
The hyperbolic characteristic speeds of Equation \ref{eq:conservative form} are the two non-zero eigenvalues $\sigma_{\text{sed},1}$ and $\sigma_{\text{sed},2}$ of the Jacobian of the flux function
\begin{equation}
  \mathbf{f}(\rho T,\rho q,\rho n_{\text{r}},\rho q_{\text{r}}) = (0, 0, v_{n_{\text{r}}} \rho n_{\text{r}}, v_{q_{\text{r}}} \rho q_{\text{r}} )^T
  \label{eq:flux function}
\end{equation}
taken with respect to the conserved quantities $\rho T$, $\rho q$, $\rho  n_{\text{r}}$, and $\rho q_{\text{r}}$.
Figure \ref{fig:speeds} shows these characteristic speeds compared to the fall speed $v_{q_{\text{r}}}$ for a representative atmospheric state.
Note that the fall speed $v_{q_{\text{r}}}$ is significantly slower than the faster characteristic speed $\sigma_{\text{sed},2}$, illustrating how using $v_{q_{\text{r}}}$ as $\tilde{v}$ can underestimate the propagation speed when choosing a time step.
We demonstrate in Section \ref{sec:sedimentation} that using $\sigma_{\text{sed},2}$ indeed results in more optimal time step choices.
For the full derivation of $\sigma_{\text{sed},1}$ and $\sigma_{\text{sed},2}$, as well as Equation \ref{eq:conservative form}, see \ref{app:hyperbolics}.

For the characteristic length scale $\tilde{z}$, the appropriate choice is perhaps less straightforward than for $\tilde{v}$.
One can select $\tilde{z}$ by considering the leading-order error term for the one-sided finite difference approximation of $x'(z)$ on a grid with spacing $\tilde{z}$, namely $\tfrac{1}{2} \tilde{z} x''(z)$.
Given that $\tfrac{1}{2} \tilde{z} x''(z)$ is a point-wise measure of absolute error for each point on a grid with spacing $\tilde{z}$, consider $\tfrac{1}{2} \tilde{z}^2 x''(z)/x(z)$ as the leading-order relative error term.
Thus, one might define a vertical grid such that $\tilde{z}^2$ is proportional to $|x(z)/x''(z)|$ to attain a particular relative error, indicating that $|x(z,t) / \frac{\partial^2 x}{\partial z^2}(z,t)|^{1/2}$, with $x$ denoting one of $n_{\text{r}}$ or $q_{\text{r}}$, is a length scale of the physical system.
Using such a length scale for $\tilde{z}$ results in a sedimentation timescale $\tilde{z}/\tilde{v}$ that can be interpreted as the timescale that must be resolved to accurately capture sedimentation in the numerical solution and is, therefore, referred to herein as the sedimentation accuracy timescale.
Another choice for $\tilde{z}$ is the grid spacing $\Delta z$ of the computational grid, which results in the timescale $\tilde{z}/\tilde{v}$ associated with the CFL condition for a hyperbolic system.
Given this timescale can be interpreted as what must be resolved for the numerical solution to be stable (i.e., remain bounded), it is herein referred to as the sedimentation stability timescale.

\begin{figure}[t!]
    \centering
    \includegraphics[width=0.99\linewidth]{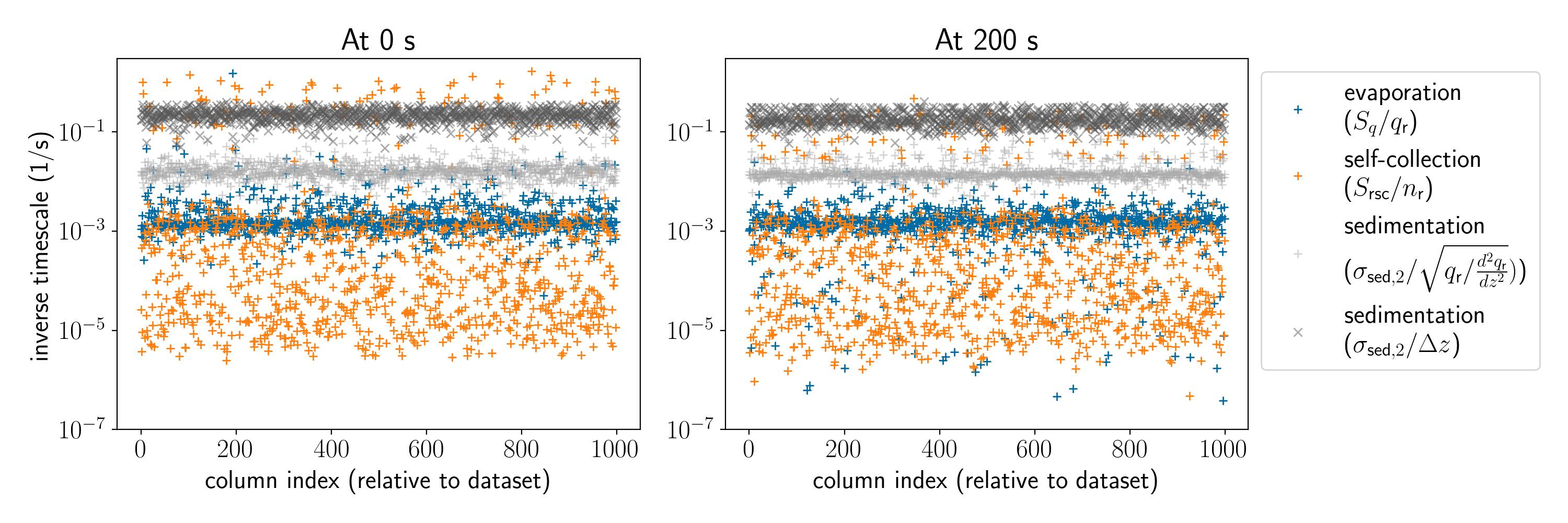}
    \caption{Inverse timescales for sedimentation, self-collection, and evaporation for 1000 columns from E3SM described in Section \ref{sec:boundary conditions} at $t=\SI{0}{\second}$ and $t=\SI{200}{\second}$. Sedimentation timescales are faster than those of the other processes for most columns, although some columns have self-collection timescales that are as fast as or faster than those of sedimentation. Additionally, the sedimentation accuracy timescale ($\sqrt{q_{\text{r}} / \frac{d^{2}q_{\text{r}}}{dz^{2}}} / \sigma_{\text{sed},2}$) is faster than the sedimentation stability timescale ($\Delta z / \sigma_{\text{sed},2}$), suggesting that sedimentation is a mildly stiff process.}
    \label{fig:process rates}
\end{figure}

With the evaporation timescale, raindrop self-collection timescale, and the two sedimentation timescales now in hand, we compute the inverse timescales at each grid level for the data described in Section \ref{sec:boundary conditions} for both $t=\SI{0}{\second}$ and $t=\SI{200}{\second}$, recording the maximum value for each inverse timescale across each column.
The $t=\SI{200}{\second}$ reference solution is obtained from a fifth-order ERK method with a fixed time step of $\Delta t=\SI{0.01}{\second}$.
Figure \ref{fig:process rates} shows that despite the traditional view of sedimentation as a \emph{fast} microphysical process that requires substepping in P3 for accuracy, the sedimentation stability timescale  is generally faster than the sedimentation accuracy timescale.
Therefore, sedimentation is perhaps more a \emph{stiff} process than a fast process, i.e., a process that restricts explicit time integration methods to smaller time steps for stability rather than to accurately capture a fast-evolving solution.
This is a significant distinction for choosing a time integration method, as stiff processes can benefit from the improved stability of implicit methods that typically allow for larger stable time steps.
Implicit methods, however, are more expensive per time step and are most beneficial when there is a large separation in the stability and accuracy timescales, which does not appear to be the case here.

Figure \ref{fig:process rates} also shows a wide spread among the self-collection, evaporation, and sedimentation accuracy timescales: some columns have sedimentation timescale for accuracy that is the fastest by an order of magnitude or more, while other columns have a raindrop self-collection timescale that is the fastest.
Whether processes can be consistently partitioned into \emph{fast} and \emph{slow} groups is another key distinction for choosing a time integration method, as such partitioned processes can benefit from being integrated by multirate methods (e.g., MRI-GARK methods discussed in \ref{sec:additive RK}).
That said, the lack of a consistently faster process in Figure \ref{fig:process rates} suggests that multirate methods may not be more competitive than single-rate methods for the rainshaft model.
This analysis of process stiffness and potential partitioning provides a blueprint for evaluating whether implicit or multirate methods could outperform explicit, single-rate methods at higher resolutions or in systems with more intrinsically stiff atmospheric processes and pronounced fast-slow processes.

%
%

\subsection{Sedimentation stability analysis} \label{sec:sedimentation}

In this subsection, we connect the stability requirements of the sedimentation process with the sedimentation stability timescale and the hyperbolic CFL condition.
Note that the backward differencing scheme used in Section \ref{sec:spatial} is equivalent to a first-order finite volume scheme for which a well-known CFL condition \cite{courant1928partiellen, leveque2002finite} exists:
\begin{align} \label{eq:cfl}
    \frac{\Delta t}{\Delta z_{i}}\max_{j} |\sigma_{\text{sed},j}| \leqslant 1.
\end{align}
Recall that $\sigma_{\text{sed},j}$ denotes an eigenvalue of the Jacobian of the flux function (Equation \ref{eq:flux function}) and is also a hyperbolic characteristic speed that appears in the sedimentation timescales.
We define the largest time step that satisfies the CFL condition (Equation \ref{eq:cfl}) as a function of $t$:
\begin{align} \label{eq:upper bound fall speed}
    \text{Stability Bound(t)} = \min_{i}\frac{\Delta z_{i}}{\max_{j} |\sigma_{\text{sed},j}(\rho_{i}, n_{\text{r},i}, q_{\text{r},i})|}.
\end{align}
This CFL stability bound is different than the one leveraged by P3 for the sedimentation substep time step:
\begin{align} \label{eq:P3 cfl}
    \text{Stability Bound}_{P3}(t) = \min_{i} \frac{\Delta z_{i}}{|v_{q_{\text{r}}}(\rho_{i}, n_{\text{r},i}, q_{\text{r},i})|}.
\end{align}
Note the P3 stability bound would be equivalent to the CFL stability bound if the fall speeds did not depend on the solution.
Regardless, neither Equation \ref{eq:upper bound fall speed} nor Equation \ref{eq:P3 cfl} will guarantee the stability of an explicit method; the integration method itself likely imposes additional stability conditions. We consider such an additional condition imposed for the forward Euler method, which assumes that the discrete sedimentation operator is sufficiently described by its Jacobian, denoted by $\bm{\mathfrak{J}}_{\text{sed}}(\bm{y}(t))$, at each time $t$:
\begin{align} \label{eq:forward Euler stability}
    \text{Upper Bound}_{\text{RK1}}(t) = \frac{2}{\max_{j} |\sigma_{\bm{\mathfrak{J}_{\text{sed}}},j}|},
\end{align}
where $\sigma_{\bm{\mathfrak{J}}_{\text{sed}}}$ are the eigenvalues of the Jacobian of the sedimentation part of $\mathfrak{F}$ in Equation \ref{eq:semidiscrete operator form}.
Given the exact computation of $\sigma_{\bm{\mathfrak{J}}_{\text{sed}}}$ can often be impractical, we will also investigate the time step chosen by an adaptive time integration method that adheres to the stability conditions using only local error estimates.
Specifically, we use a second-order ERK method from SUNDIALS that utilizes an embedded first-order method to estimate temporal error and adapt the time step.
The second-order ERK method has the same linear stability bound described in Equation \ref{eq:forward Euler stability} for operators with purely real eigenvalues.

\begin{figure}[t!]
    \centering
    \includegraphics[width=0.7\textwidth]{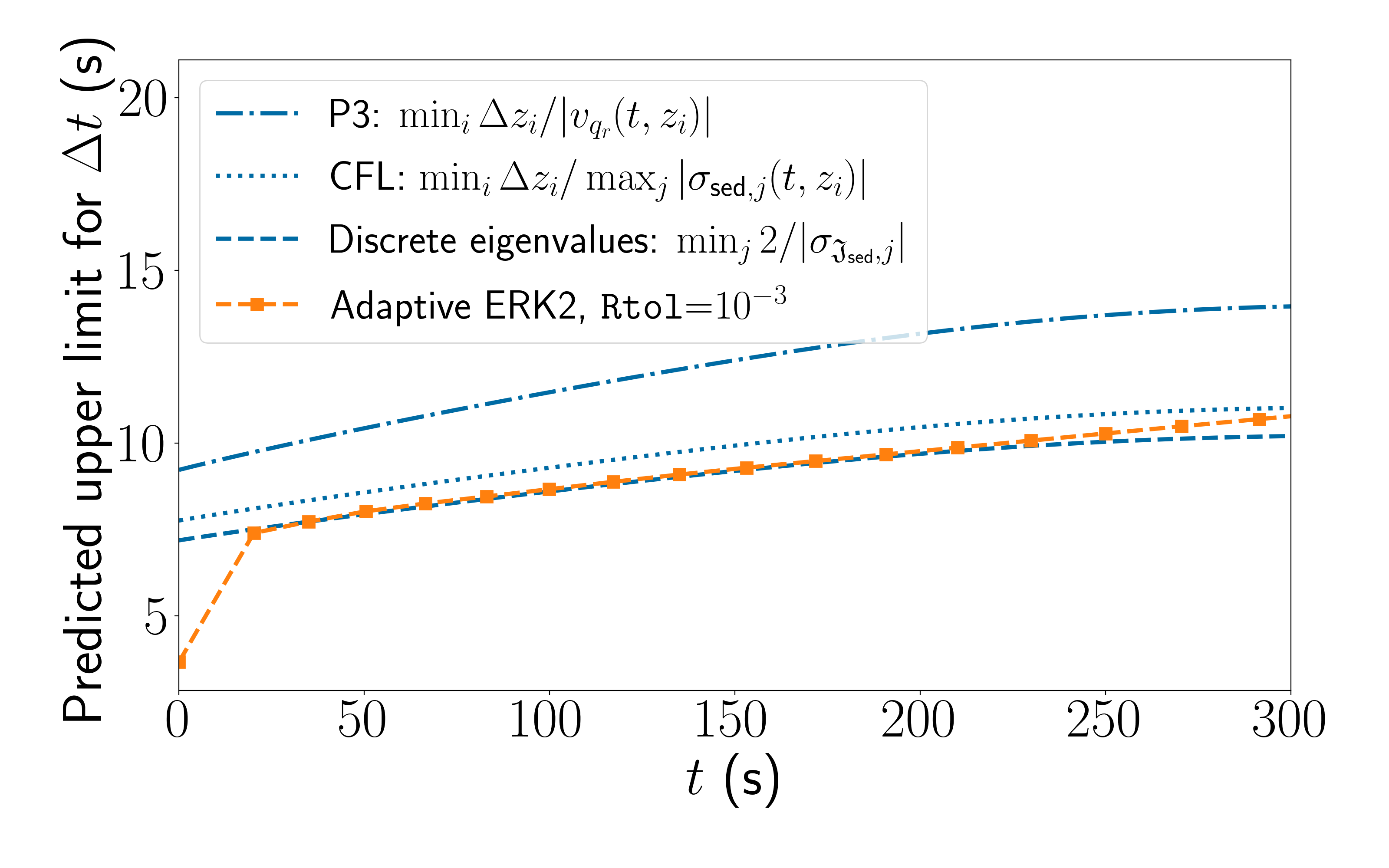}
    \caption{Predicted stability bound for column 37, with sedimentation as the only active process, using (i) the P3 condition using fall speeds (see Equation \ref{eq:P3 cfl}) (ii) the CFL condition (see Equation \ref{eq:cfl}), and (iii) the forward Euler method requirement on discrete eigenvalues when considering the rainshaft model (Equation \ref{eq:rainshaft model}). We compare these bounds with the time step selected by SUNDIALS with a second-order ERK method which uses an embedded first-order method to estimate temporal error and adapt the time step. The SUNDIALS controller performs well in correctly capturing the CFL condition (Equation \ref{eq:cfl}).}
    \label{fig:cfl sedimentation}
\end{figure}

Figure \ref{fig:cfl sedimentation} shows the CFL stability bound (Equation \ref{eq:upper bound fall speed}), the P3 stability bound (Equation \ref{eq:P3 cfl}), the forward Euler stability bound (Equation \ref{eq:forward Euler stability}) for a particular column of the reference solution used for the timescales in Section \ref{sec:timescales}, as well as the time step size chosen by the adaptive second-order ERK method.
Note the P3 stability bound overestimates the stable time step specified by the CFL stability bound, which is consistent with the significant difference between the fall speed $v_{q_{\text{r}}}$ and the hyperbolic characteristic speeds $\sigma_{\text{sed},j}$ shown in Figure \ref{fig:speeds}.
Such an overestimation can result in the P3 stability bound producing unstable solutions that require limiters (such as the ones leveraged by P3) or other remediation to remove the instabilities.
Granted the CFL stability condition also overestimates the forward Euler stability bound, that overestimation is less severe than that of the P3 stability bound.
Lastly, we note the time step selected by an adaptive scheme in this scenario approximates the stability condition of the forward Euler method remarkably well.
More details of the methods with adaptive time steps will be introduced in Section \ref{sec:additive RK}.

%
%

\subsection{Self-collection and evaporation stability} \label{sec:evaporation and self-collection}

We now discuss the overall stability requirements when evaporation and self-collection processes are included with sedimentation.
Assuming the behavior of the resulting discrete nonlinear operator is sufficiently described by its Jacobian, denoted $\bm{\mathfrak{J}}(\bm{y}(t))$, at each time $t$, we can again use linear stability analysis to express the stability requirements on time step size in terms of the eigenvalues of $\bm{\mathfrak{J}}$.
Linear stability analysis requires the eigenvalues to fall within the stability region prescribed by the particular time integration method being used.
As an example, the forward Euler stability condition (Equation \ref{eq:forward Euler stability}) requires the eigenvalues to fall within a unit circle centered around $z=-1$ in the complex plane.
Here, we will consider the stability regions prescribed by particular second-, third-, and fourth-order ERK methods.

\begin{figure}[t!]
    \centering
%
    \centering
    \includegraphics[width=0.99\textwidth]{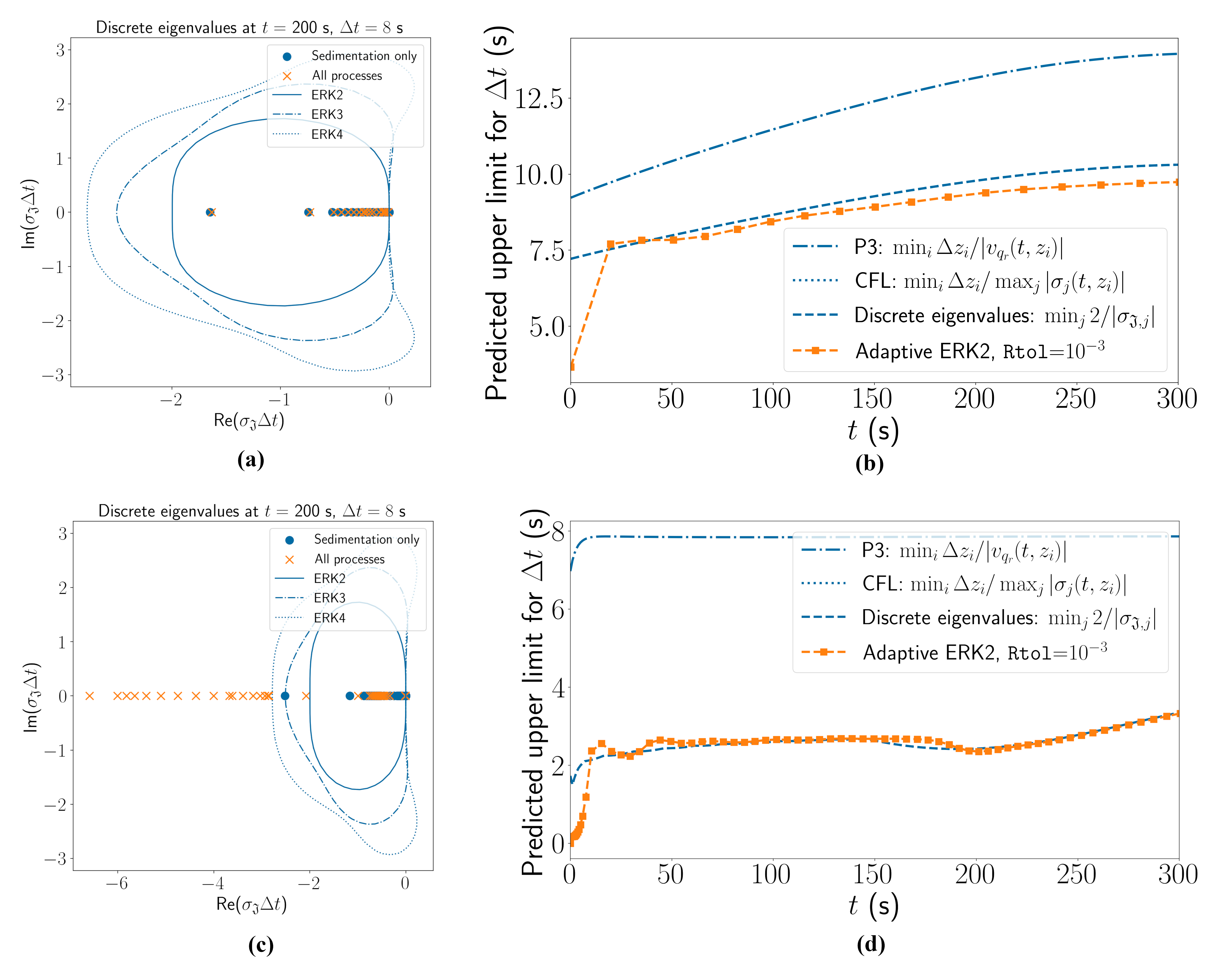}
    \caption{(a) Eigenvalues of the Jacobian of the discrete sedimentation operator (blue circle) and of the overall operator $\bm{\mathfrak{F}}$ (orange x) at $t=\SI{200}{\second}$ for column 37. Stability regions for second, third, and fourth-order ERK methods are overlaid on the eigenvalues. (b) The predicted stability requirement on time step based on the P3 condition (i.e., Equation \ref{eq:P3 cfl}, dashed-dotted line), the CFL condition (i.e., Equation \ref{eq:cfl}, dotted line), and the Forward Euler method requirement on discrete eigenvalues (i.e., Equation \ref{eq:forward Euler stability}, dashed line), and the time step chosen by a second-order ERK method with adaptive time step in SUNDIALS (dashed orange line) with a relative tolerance of $10^{-3}$. (c) and (d) The analogous plots for column 821.}
    \label{fig:stability conditions columns 5 and 37}
\end{figure}

Figure \ref{fig:stability conditions columns 5 and 37} shows the stability regions and computed eigenvalues of $\bm{\mathfrak{J}}$ at each time $t$ for both column 37, in which the sedimentation stability timescale is faster than other timescales, and column 821, in which other timescales are faster than the sedimentation stability timescale.
The eigenvalues of the Jacobian of the discrete sedimentation operator are also shown for comparison with the stability requirements for sedimentation discussed in the prior section, as well as the CFL stability bound, the P3 stability bound, the forward Euler stability bound, and the time step size chosen by the adaptive second-order ERK method.
Note that the eigenvalues for column 37 show a negligible shift from those of the discrete sedimentation operator, because the sedimentation stability timescale is faster than other timescales.
However, for column 821, self-collection and evaporation significantly shift the eigenvalues of the discrete sedimentation operator, because the sedimentation stability timescale is slower than other timescales.
For both columns we observe that the adaptive step method is able to correctly identify and approximately adhere to the forward Euler stability condition (Equation \ref{eq:forward Euler stability}).
Such an observation motivates the utility of using a method with adaptive time steps based on error-estimation, rather than time steps determined by either a CFL condition, which can allow for unstable time step sizes, or by linear stability analysis, which can require a large or infeasible amount of computational expense for the eigenvalues.

%
%
\section{Evaluation of Time Integration Methods} \label{sec:time integration}

We evaluate a number of higher-order time integration methods applied to the rainshaft model in Section \ref{sec:rainshaft model} with the aim of improving both the accuracy and efficiency of the original P3 time integration method described in Section \ref{sec:P3}. We consider ERK, DIRK, ImEx, operator splitting, and MRI-GARK methods. Brief overviews of these methods are provided in the subsequent subsections. A discussion of the efficiency of each method may be found in its respective subsections and, where relevant, we explain how the process analysis in Section \ref{sec:process analysis} may be used to anticipate method efficiency and guide selection of the best-performing methods.

We compare the performance of each method over the $N_{\text{col}}$ columns of P3 described in Section \ref{sec:boundary conditions}. The precise metric we use is a weighted root mean square (WRMS) norm, which allows us to reasonably aggregate results over multiple columns into a single metric and is defined as follows:
\begin{equation} \label{eq:weighted rms}
    ||\cdot||_{\text{WRMS}} := \sqrt{\frac{1}{N_{\text{col}}} \sum_{i=1}^{N_{\text{col}}} f_{\text{r},i}\sum_{j=1}^{N_{z,i}} \frac{(\Delta z)_{ij}}{z_0} \left[ (\cdot)_{ij} \right]^{2} }.
\end{equation}
Here, $(\cdot)_{ij}$ denotes the value of the argument at the $j$th grid point of the $i$th column at time $t$. This norm may be viewed as the average $\ell^2$ norm over the column height weighted by the precipitation fraction $f_{\text{r},i}$. We focus our presentation of numerical results on the rain mass mixing ratio $q_\text{r}$, which, in our observation, is the prognostic variable typically containing most of the temporal approximation error. The results for the other prognostic variables (not shown) are similar enough to those for $q_{\text{r}}$ that they would lead to the same recommendations in any case.

Since a closed form solution to Equation \ref{eq:rainshaft model} is not available, the error due to the choice of time integration method at time $t$ is estimated by comparing the difference between $q_{\text{r}}$ obtained using one of the time integration methods described below with a highly-resolved reference solution $q_{\text{r,ref}}$ obtained with a third-order ERK method using adaptive time stepping guided by relative tolerance $\texttt{Rtol} = 10^{-13}$ and absolute tolerances of \num{1.e-6}, \num{1.e-8}, \num{1.e-9}, and \num{1.e-17} for $T$, $q$, $n_{\text{r}}$, and $q_{\text{r}}$, respectively. These tolerances are chosen to yield a highly accurate reference solution (orders of magnitude more accurate than any practical application needs). The WRMS error estimate, herein referred to as WRMSE, is then given by $||q_{\text{r}}(t) - q_{\text{r,ref}}(t)||_{\text{WRMS}}$. We choose to evaluate the WRMSE at $t=\SI{300}{\second}$ which is the microphysics time step used in E3SM.

We are interested not only in the temporal discretization error estimate associated with each of the time integration methods but also their efficiency. To this end, we compare the WRMSE against the total wall clock time required to integrate all $N_{\text{col}}$ columns to the final time $t = \SI{300}{\second}$. These results are provided in the form of work-precision diagrams which show how the estimated discretization error decreases as the wall clock time is increased, typically due to decreasing the time step for methods with fixed time steps or the relative error tolerance for methods with adaptive time steps. 

Additionally, we also refer to the relative WRMSE in the following sections, which is defined as
\begin{equation}
    \text{Relative WRMSE} := \frac{||q_{\text{r}}(t) - q_{\text{r,ref}}(t)||_{\text{WRMS}}}{||q_{\text{r,ref}}(t)||_{\text{WRMS}}} \times 100\%.
\end{equation}

\subsection{Operator Splitting Methods} \label{sec:operator splitting}


As the current P3 method is based on a Lie--Trotter operator splitting method, it is natural to consider higher-order operator splitting schemes such as a Strang splitting \cite{strang1968construction}. These schemes still integrate local microphysical processes separately from sedimentation but may evaluate processes repeatedly (i.e., using multiple stages per step). See \ref{app:time integration methods} for additional details on the particular operator splitting methods evaluated.
Notably, at order three, a splitting method with real coefficients must perform integrations backwards in time \cite{blanes2005necessity} which can potentially degrade stability and efficiency. We first provide operator splitting results using the same substepping procedure for sedimentation as in P3 and employing the same limiters to prevent negative solution values. The impact of the limiters on solution accuracy and convergence is examined more closely in Section \ref{sec:limiters}.

\begin{figure}
    \centering
    \includegraphics[width=0.99\linewidth]{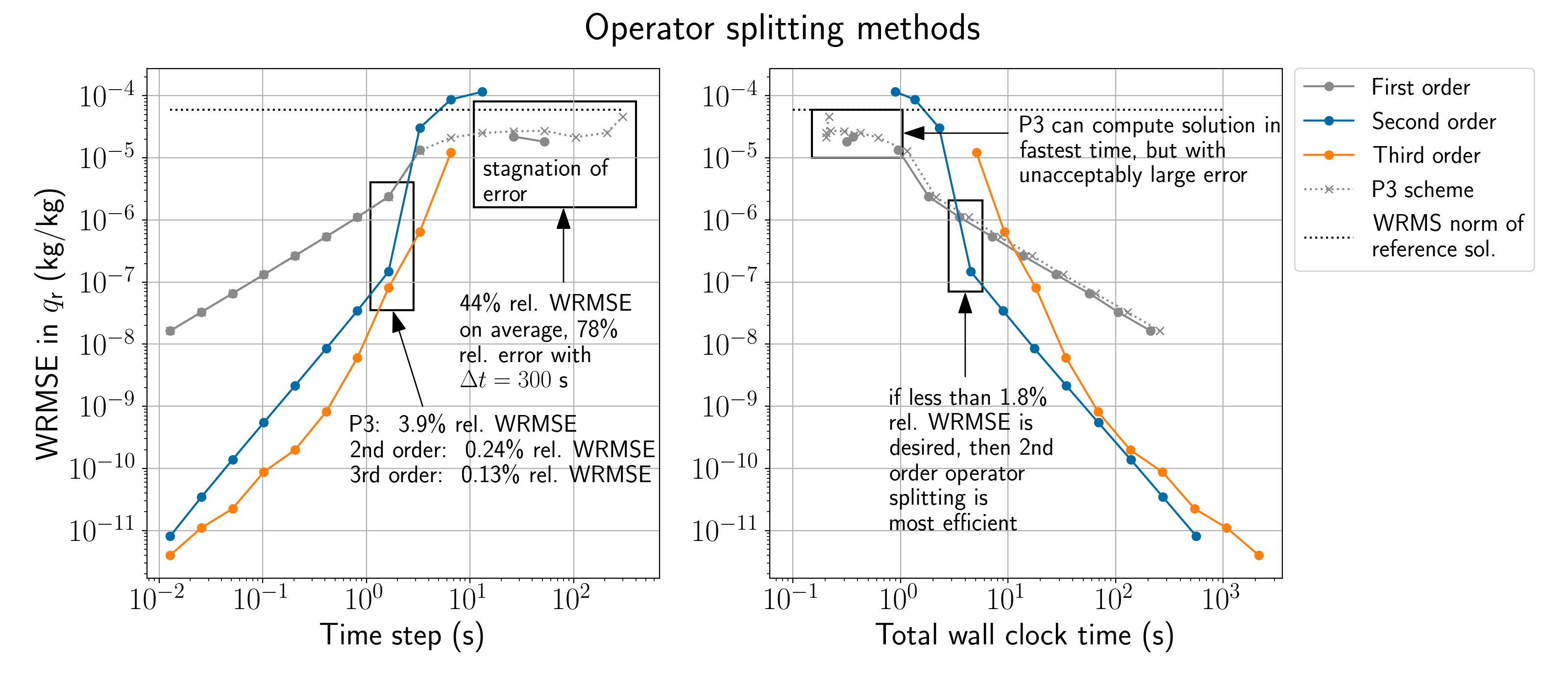}
    \caption{(Left) WRMSE of $q_{\text{r}}$ vs. local process time step when various operator splitting methods in SUNDIALS are applied to the rainshaft model. (Right) WRMSE of $q_{\text{r}}$ vs. total wall clock time to integrate all columns to $t=\SI{300}{\second}$. In both plots, the WRMS norm of the reference solution is provided by the dotted black line in order to contextualize the relative magnitude of the error estimates attained by each method. When the WRMSE of the solution provided by a particular method is equal to that of the WRMS norm of the reference solution, the relative WRMSE of that solution is 100\%.}
    \label{fig:opsplit regularized all columns}
\end{figure}

Figure \ref{fig:opsplit regularized all columns} shows the convergence behavior (the WRMSE in $q_{\text{r}}$ vs. the local process time step) of first, second, and third-order operator splitting methods compared to P3's time integration method. For time steps less than approximately \SI{3}{\second}, the second and third-order operator splitting methods attain significantly smaller (between one and three full orders of magnitude) WRMSE than P3. For larger time steps, the first and third-order methods became unstable. The P3 method is able to take time steps up to \SI{300}{\second} (i.e., the current time step in E3SM), but we reiterate the point from Section \ref{sec:P3 accuracy} that solutions at such large time steps have extremely large error. In particular, for time steps between $\sim\SI{3}{\second}$ and \SI{300}{\second}, the relative WRMSE of the P3 scheme is approximately \num{44}\%. At the default time step of \SI{300}{\second}, the relative WRMSE is \num{78}\%.

We remark that the third-order method exhibits a degraded convergence rate caused by a discontinuity in the derivative of the rain self-collection tendency. While regularization of this tendency restores third-order convergence for this method, unlike the other regularizations described in Section \ref{sec:regularization}, its effect on the physics is large enough that the regularized scheme converges to a noticeably different solution than the original un-regularized scheme. Furthermore, even with that regularization, the third-order splitting method is only the most efficient when an estimated WRMSE of \SI{1.e-9}{\kilogram\per\kilogram} or smaller is desired (not shown). As a result, we have decided not to use the self-collection regularization in this study, though a brief description of this regularization is included in supplemental information for the curious reader.

When the WRMSE of the numerical solution obtained with operator splitting methods or the P3 scheme is less than $\sim$\SI{1.e-6}{\kilogram\per\kilogram} (corresponding to a relative difference of ~\num{1.8}\%), then the second-order operator splitting method is most efficient. For WRMSE larger than $\sim$\SI{1.e-6}{\kilogram\per\kilogram}, the P3 scheme and related first-order operator splitting are still the most efficient, and none of the operator splitting methods improve on the wall clock time to break out of the stagnant convergence region of the P3 scheme. Additionally, it is notable that although the third-order operator splitting method provides the smallest WRMSE of $q_{\text{r}}$ at a given time step, it is \emph{not} the most efficient of the methods surveyed here due to the cost of the additional stages.

\subsubsection{Effect of size and positivity limiters on convergence and solution accuracy} \label{sec:limiters}

It is worthwhile to further discuss the large WRMSE associated with large time steps in P3 before analyzing the behavior of other time integration methods. It is clear from the left panel of Figure \ref{fig:opsplit regularized all columns} that both P3 and the first-order operator splitting methods achieve first-order convergence when the time step is less than $\sim\SI{3}{\second}$. At larger time steps outside of the asymptotic regime, the convergence stagnates; for instance, when the time step is reduced from \SI{300}{\second} to \SI{3}{\second}, the estimated WRMSE of the P3 solution is not predictably changed. 

\begin{figure}
    \centering
    \includegraphics[width=0.95\linewidth]{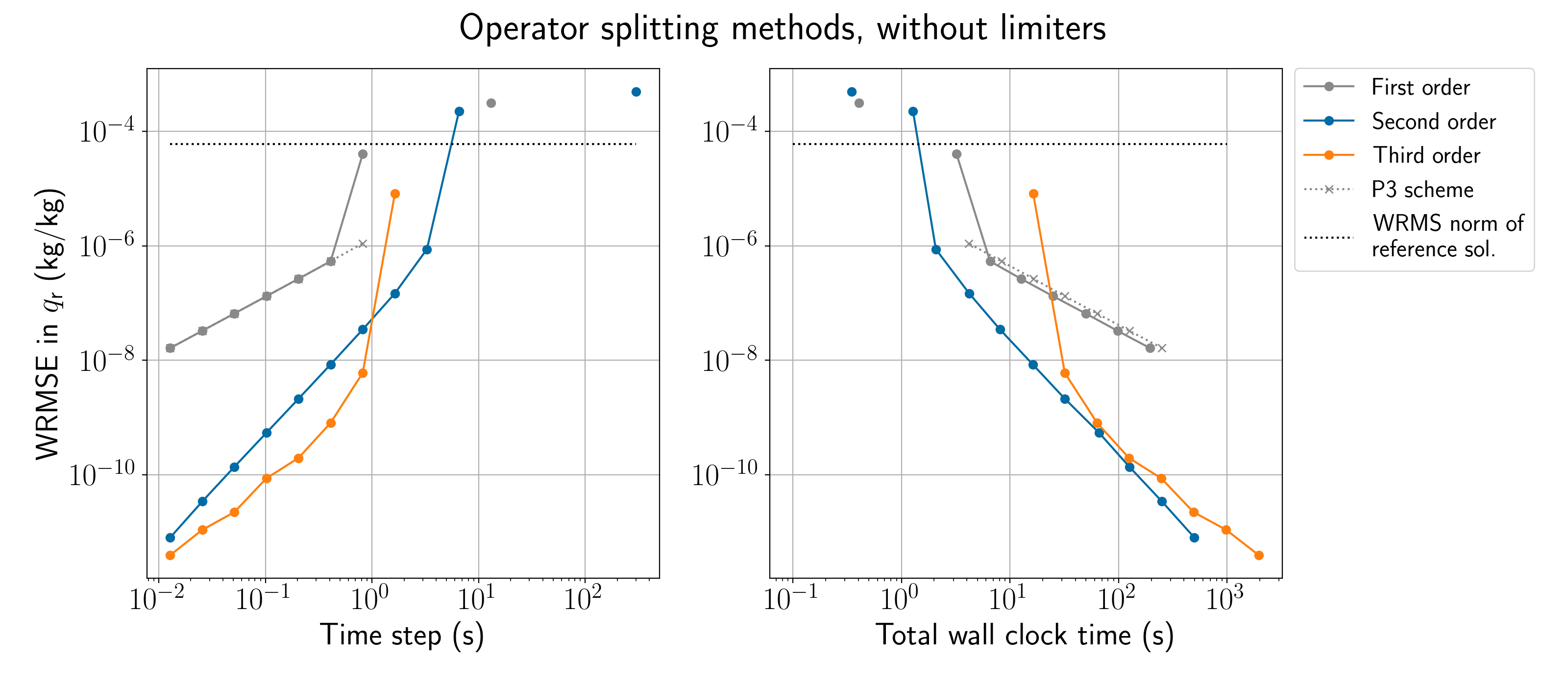}
    \caption{(Left) WRMSE vs. local process time step when various operator splitting methods in SUNDIALS are applied to the rainshaft model with positivity and size limiters turned off for all methods. (Right) WRMSE vs. total wall clock time to integrate all columns to $t=\SI{300}{\second}$. In both panels, omitted points correspond to time steps for which the model aborted or produced NaN when no limiters were active.}
    \label{fig:opsplit regularized no limiter}
\end{figure}

This stagnated convergence of the numerical solution is due to the positivity and size limiters in P3. To see this, we run the same simulations as in Figure \ref{fig:opsplit regularized all columns} except with the positivity and size limiters disabled. The largest time step each method can take while maintaining a WRMSE estimate below \num{100}\% without limiters active is roughly between \SI{1}{\second} and \SI{7}{\second} as observed in Figure \ref{fig:opsplit regularized no limiter}. These results serve as a cautionary tale on the extent to which limiters can hide the inaccuracy of numerical solutions. Model ``crashes'' due to instability can be costly, for instance if they delay urgent weather forecasts, or force parts of expensive climate simulations to be repeated. Therefore it is understandable that parameterization developers rely on limiters to prevent the model state from becoming too unrealistic and potentially crashing or otherwise ruining a simulation, particularly when such limiters are expected to have a very small effect on the solution (or to trigger exceedingly rarely). Unfortunately, this approach to ensuring model robustness can lead to a situation where users are never notified about large inaccuracies in a parameterization, even if these errors occur regularly and not only in rare edge cases.
As a result, an additional goal in our subsequent investigations of time integration methods is to ensure that the time integration is accurate enough to always result in an output state that is reasonable enough to avoid model crashes, even in the absence of limiters.

\subsection{Runge--Kutta-Based Methods} \label{sec:additive RK}

To improve upon the performance benefits observed with second-order operator splitting methods, we now consider the class of additive Runge--Kutta (ARK) methods \cite{cooper1980additive,ascher1997implicit} which are based on a partitioning of Equation \ref{eq:semidiscrete operator form} of the form
\begin{equation} \label{eq:ImEx ODE}
    \frac{\partial\bm{y}(t)}{\partial t} = \bm{\mathfrak{F}}(\bm{y}(t)) = \bm{\mathfrak{F}}^E(\bm{y}(t)) + \bm{\mathfrak{F}}^I(\bm{y}(t)).
\end{equation}
Here, $\bm{\mathfrak{F}}^E$ denotes a non-stiff RHS function which will be treated explicitly, and $\bm{\mathfrak{F}}^I$ denotes a stiff RHS function which will be treated in a diagonally implicit manner.
We consider three ways to partition sedimentation, self-collection, and evaporation into the two functions.
First, based on the analysis of Section \ref{sec:process analysis}, it is natural to treat sedimentation implicitly and the other two processes explicitly, resulting in an ImEx method.
Second, treating all processes explicitly (i.e., $\bm{\mathfrak{F}}^E$ encompasses all processes) results in an ERK method.
Finally, treating all processes in a diagonally implicit manner (i.e., $\bm{\mathfrak{F}}^I$ encompasses all processes) results in a DIRK method.

A key motivating factor in the exploration of ARK methods in this work is the desire to circumvent the restrictive sedimentation substepping procedure utilized in both P3 and the operator splitting methods explored in Section \ref{sec:operator splitting}. In particular, the more favorable stability properties of DIRK and ImEx methods compared to their explicit counterparts indicate that these methods may be more efficient than the methods surveyed thus far. Although time integration schemes for sedimentation which have some degree of implicitness have been explored and found useful in some models \cite{guo2021sedimentation,gettelman2023pumas}, they typically use fall speeds calculated from the atmospheric state at the current (rather than next) time step, to avoid solving a nonlinear system of equations. Such an approximation is valid when the fall speeds do not change too dramatically in a single time step. In contrast, we use DIRK and ImEx methods where all quantities in the implicit terms are evaluated at the same Runge--Kutta stage, increasing the consistency and generality of the methods at the cost of requiring a nonlinear solver to be run at each time step.

Another time integration approach is to utilize the class of MRI-GARK methods \cite{sandu2019class}, which allow partitioning of the microphysical processes such that ``faster'' processes are integrated with a smaller time step than other processes.
There are two key advantages of such methods over the simpler operator splitting methods.
First, MRI-GARK methods tend to be more accurate due to their tighter coupling.
While operator splitting only couples partitions through initial conditions, MRI-GARK also employs a polynomial forcing function to approximate slow dynamics on the finer time grid used to resolve the fast dynamics.
Second, high-order MRI-GARK methods do not require backwards integrations to achieve high order accuracy. The conventional assumption that sedimentation is a fast process and self-collection and evaporation are slow processes, which has motivated use of the current P3 scheme, would also suggest higher order operator splitting methods and MRI-GARK methods to be competitive alternatives to the P3 scheme. However, as shown in the timescale analysis of Section \ref{sec:process analysis}, the process timescales are not sufficiently separated across all columns to motivate use of these methods. Indeed, we show in the subsequent sections that the operator splitting methods covered in Section \ref{sec:operator splitting} and MRI-GARK methods perform worse for this model than explicit methods which treat all processes with the same time step. More information about process partitioning for each method may be found in Section \ref{sec:implementation}, and references to all time integration methods applied to the rainshaft model may be found in \ref{app:time integration methods}.


\subsubsection{Time step adaptivity}

As noted in Sections \ref{sec:sedimentation} and \ref{sec:evaporation and self-collection}, time step adaptivity is able to more accurately estimate the maximum stable time step at a given time than the P3 substepping procedure based on sedimentation fall speeds. As such, our evaluation of ARK and MRI-GARK methods in the following sections include results for both fixed and adaptive time stepping. The adaptive time stepping we employ follows the approach outlined in \cite[Section II.4]{hairer1993solving} and \cite[Section 3.1]{reynolds2023arkode}.
The numerical solution $\bm{y}_{n+1} \approx \bm{y}(t_{n+1})$ of a method is compared to an embedded solution $\widehat{\bm{y}}_{n+1}$, which is one order of accuracy lower, using the following error estimate:
%
\begin{equation} \label{eq:local_error_estimate}
    \epsilon_n = \sqrt{\frac{1}{d} \sum_{i = 1}^{d} \left( \frac{\bm{y}_{n+1,i} - \widehat{\bm{y}}_{n+1,i}}{\texttt{Atol}_i + \texttt{Rtol} \left| \bm{y}_{n,i} \right|} \right)^2}.
\end{equation}
Here, $\texttt{Atol}_i$ and $\texttt{Rtol}$ are user-specified absolute and relative tolerances and $d$ is the size of $\bm{y}_{n+1}$. If $\epsilon_{n} \leqslant 1$, the time step is deemed acceptable, and if $\epsilon_{n} > 1$, it is rejected, and the time step is reduced.

\subsubsection{Implementation in SPAECIES} \label{sec:implementation}

In this section, we describe key details of the SPAECIES software implementation and hyperparameters for the time integrators that ensure a reliable and accurate solution over a wide range of initial conditions. SPAECIES uses the ARKODE package of the SUNDIALS library \cite{hindmarsh2005sundials,reynolds2023arkode} to provide implementations of the integrators discussed in Sections \ref{sec:operator splitting}-\ref{sec:additive RK}.
Details on ARK and MRI-GARK implementations can be found in \cite[Section 2.2-2.3]{reynolds2023arkode}, while operator splitting details can be found in \cite[Section 2.3]{roberts2025new}.
We summarize the particular schemes used in Table \ref{tab:methods} which correspond to the default methods in ARKODE using SUNDIALS version 7.3.0.
For the MRI-GARK and operator splitting methods, there are inner ODEs that must be solved, and corresponding numerical integrators must be selected for these to fully specify the numerical method.
We use the default ERK method of the same order as the MRI-GARK or operator splitting method.

For the implicit integrators, we use a standard, modified Newton's method \cite[Section IV.8]{hairer1996solving} to solve the nonlinear systems arising from the stage equations.
For the DIRK methods, Newton's method requires (an approximation of) the Jacobian $\partial \bm{\mathfrak{F}}/\partial \bm{y}$, and for the ImEx methods, this requires (an approximation of) $\partial \bm{\mathfrak{F}}^I/\partial \bm{y}$.
SPAECIES implements analytic Jacobians for all RHS functions and the intermediate functions used within them.
Each Newton iteration requires the solution of a linear system of equations.
Due to the small dimension of these systems (the largest of which has size \num{80}), we find that a dense LAPACK linear solver performs well.

As is typical with adaptive methods, selecting appropriate absolute tolerances in Equation \ref{eq:local_error_estimate} is critical to ensuring solution positivity while balancing accuracy with the cost of resolving quantities close to zero.
In the rainshaft application, if absolute tolerances are selected too small, a few problematic initial and boundary conditions cause the integrators to take unnecessarily small time steps without physically meaningful improvements to the accuracy.
These tolerances and other method details are summarized in Table \ref{tab:integrator settings}.


\begin{table}[ht!]
    \centering
    \begin{tabular}{p{1.5cm}|p{3.5cm}|p{1.5cm}|p{1.5cm}|p{1.3cm}|p{2cm}}
        Method & Partitioning & Nonlinear solver & Linear solver & Limiters & Abs. tols. $(T, q, n_{\text{r}}, q_{\text{r}})$ \\ \hline
        P3 & Partition 1: self-collection, evaporation \newline Partition 2: sedimentation (substepped) & --- & --- & Yes & --- \\ \hline 
        Operator splitting & Partition 1: self-collection, evaporation \newline Partition 2: sedimentation (substepped) & --- & --- & Yes & --- \\ \hline 
        Explicit Runge--Kutta & --- & --- & --- & No & (\num{1.e-6}, \num{1.e-8}, \num{1.e-9}, \num{1.e-17}) \\ \hline 
        Diagonally Implicit Runge--Kutta & --- & Newton's Method & Dense LAPACK & No & (\num{1.e-6}, \num{1.e-8}, \num{1.e-9}, \num{1.e-17}) \\ \hline 
        Implicit-Explicit Runge--Kutta & Implicit: sedimentation \newline Explicit: self-collection, evaporation & Newton's Method & Dense LAPACK & No & (\num{1.e-6}, \num{1.e-8}, \num{1.e-9}, \num{1.e-17}) \\ \hline 
        MRI-GARK & Fast: sedimentation \newline Slow: self-collection, evaporation & --- & --- & No & (\num{1.e-6}, \num{1.e-8}, \num{1.e-9}, \num{1.e-17})
    \end{tabular}
    \caption{Integrator settings for the methods in Sections \ref{sec:fixed step results} and  \ref{sec:adaptive step results}. Cells with --- indicate not applicable.
    }
    \label{tab:integrator settings}
\end{table}

\subsubsection{Results with fixed time steps} \label{sec:fixed step results}

The disadvantage of methods with fixed time steps is that one must choose the smallest time step over the course of the entire simulation that achieves both stability and the desired accuracy. Typically, this time step varies significantly as $t$ increases. For instance, previous results in Section \ref{sec:process analysis} (Figure \ref{fig:stability conditions columns 5 and 37}, dashed blue line in panel (b)) demonstrate that when $t$ is close to \SI{0}{\second}, the maximum time step for stability can be nearly \num{1.4}x smaller than it is at $t = \SI{300}{\second}$. Adaptive time step controllers, like those in SUNDIALS, allow one to account for these variations in maximum time step. Moreover, methods which use a nonlinear solver can respond to a convergence failure by reducing the time step which is not possible with fixed time steps. Nevertheless, we first evaluate ARK methods with fixed time steps in order to set a baseline for adaptive time stepping. These results allow us to verify that correct theoretical convergence rates are achieved, provide insights into the largest stable time step each method can take, and motivate the use of adaptive time stepping to improve computational efficiency.

\begin{figure}[tb]
    \centering
    \includegraphics[width=0.99\linewidth]{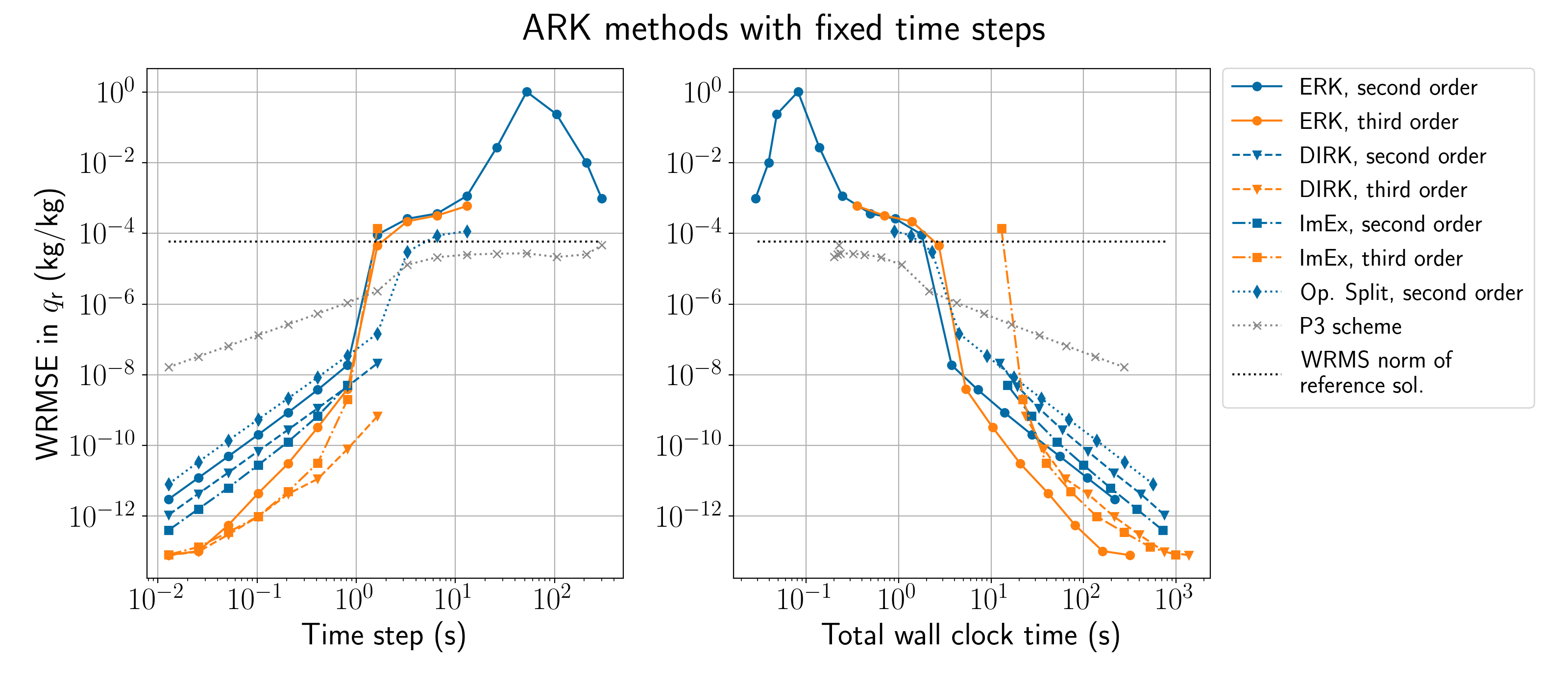}
    \caption{(Left) WRMSE vs. time step for explicit Runge--Kutta (ERK), diagonally implicit Runge--Kutta (DIRK), and implicit-explicit (ImEx) methods with fixed time steps. The result from the second-order operator splitting method from Section \ref{sec:operator splitting} is provided for comparison. (Right) WRMSE vs. total simulation wall clock time.}
    \label{fig:ark fixed step all columns}
\end{figure}

We first consider fixed time steps between \SI{0.0128}{\second} and \SI{300}{\second}. The WRMSE for solutions with these varying time steps and their corresponding total simulation wall clock times are provided in Figure \ref{fig:ark fixed step all columns}. Explicit methods can take time steps as large as $\sim$\num{20}--\SI{200}{\second}, though the WRMSE is extremely large in these cases, often far exceeding \num{100}\% relative error. Explicit methods that treat all processes with the same time step are more efficient than the operator splitting methods explored in Section \ref{sec:operator splitting}. For instance, the second-order explicit method produces a WRMSE about an order of magnitude smaller than that of the second-order operator splitting method at comparable time steps and can achieve a target WRMSE roughly five times faster as well. DIRK and ImEx methods return solutions with lower WRMSE than their explicit counterparts at comparable time steps, but the efficiency results show this improvement is more than offset by the higher computational cost of solving the nonlinear stage equations. Despite the promise of being able to circumvent the stability requirements of the explicit methods, the largest stable time steps for DIRK and ImEx methods are in the range of $\sim$\num{1}--\SI{2}{\second}. Significantly larger time steps could not be taken even with positivity and size limiters applied. These small time steps result from convergence failures and nonphysical values that arise within the nonlinear solve. In principle, limiters could be added within the nonlinear solve as well, but we do not expect such limiting to make up for the performance gap with explicit schemes.

In general, we hypothesize that the weakly stiff nature of the sedimentation process and the relatively large size of the process timescales for accuracy are the causes of the overall poor performance of DIRK and ImEx schemes here. While it is true that the sedimentation timescale for accuracy is longer than the corresponding timescale for stability (see Figure \ref{fig:process rates}), the latter is not short enough relative to the timescales of the other processes to offset the cost of the nonlinear stage solves in the DIRK and ImEx methods. Nevertheless, the analysis of sedimentation timescales provides valuable insights into how practitioners can decide whether DIRK and ImEx schemes are suitable for a particular process.

We note that both third-order DIRK and ImEx methods exhibit degraded convergence rates when small time steps are taken. As with the case of the third-order operator splitting method, this degraded convergence is largely caused by insufficient smoothness of the rain self-collection tendency. Even when the regularization of self-collection is employed (not shown), the third-order DIRK and ImEx methods are never the most efficient in achieving a given error level.

The results with fixed time steps presented here suggest that ERK methods are promising, although reducing the WRMSE of the P3 scheme below what the current method produces at a \SI{300}{\second} time step still requires significantly higher wall clock time. We address this efficiency gap with adaptive time stepping in Section \ref{sec:adaptive step results}.    

\subsubsection{Results with adaptive time stepping} \label{sec:adaptive step results}

\begin{figure}[bt]
    \centering
    \includegraphics[width=0.80\linewidth]{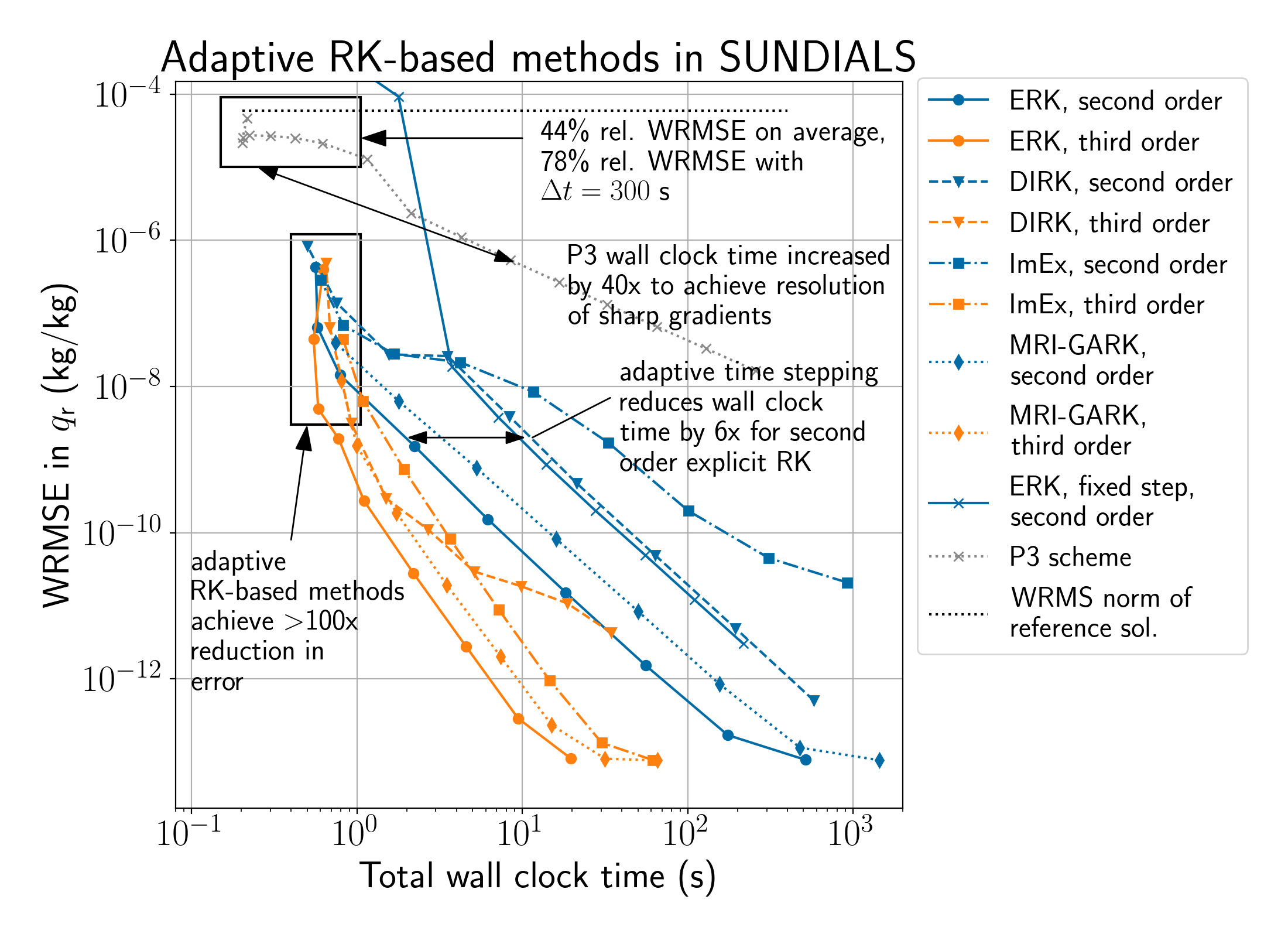}
    \caption{WRMSE vs. total simulation wall clock time for ARK (including explicit Runge--Kutta (ERK), diagonally implicit Runge--Kutta (DIRK), and implicit-explicit (ImEx))  and MRI-GARK methods with adaptive time stepping. The result from the second-order explicit method with fixed time steps from Section \ref{sec:fixed step results} is provided as a comparison.}
    \label{fig:ark adaptive step all columns}
\end{figure}

We consider ERK, DIRK, ImEx, and MRI-GARK methods and vary the relative tolerance in Equation \ref{eq:local_error_estimate} to achieve different levels of accuracy. Figure \ref{fig:ark adaptive step all columns} shows the WRMSE with corresponding total simulation wall clock times for relative tolerances between \num{1.e-2} and \num{1.e-11}; absolute tolerances are provided in Table \ref{tab:integrator settings}. The fastest wall clock times are achieved at the largest relative tolerance of \num{1.e-2}. Specifically, the second-order ERK method achieves a roughly \num{100}x reduction in WRMSE over the P3 scheme (at its default \SI{300}{\second} time step) while only modestly increasing wall clock time by a factor of \num{2.6}. Adaptivity of the time steps is crucial to achieving this result and provides roughly a $\num{6}$x reduction in wall clock time over the same method with fixed time steps. 

\begin{figure}[htb]
    \centering
    \includegraphics[width=0.85\linewidth]{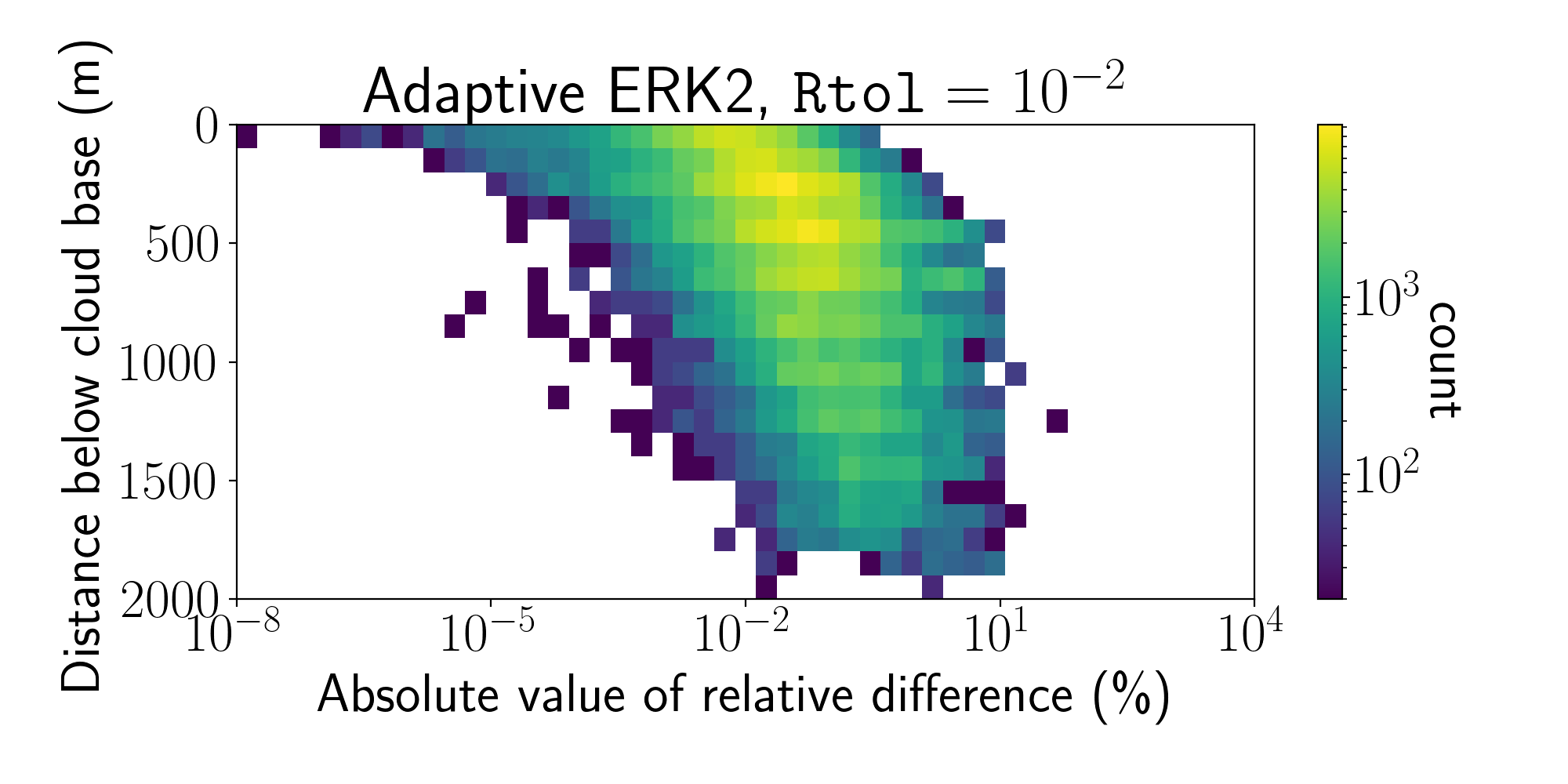}
    \caption{2D histogram of the absolute value of the pointwise error estimate of the second-order ERK (ERK2) solution with adaptive time step and $\texttt{Rtol} = 10^{-2}$ as a percentage relative to the reference solution.}
    \label{fig:SUNDIALS histogram}
\end{figure}

\begin{figure}[htb]
    \centering
    \includegraphics[width=0.99\linewidth]{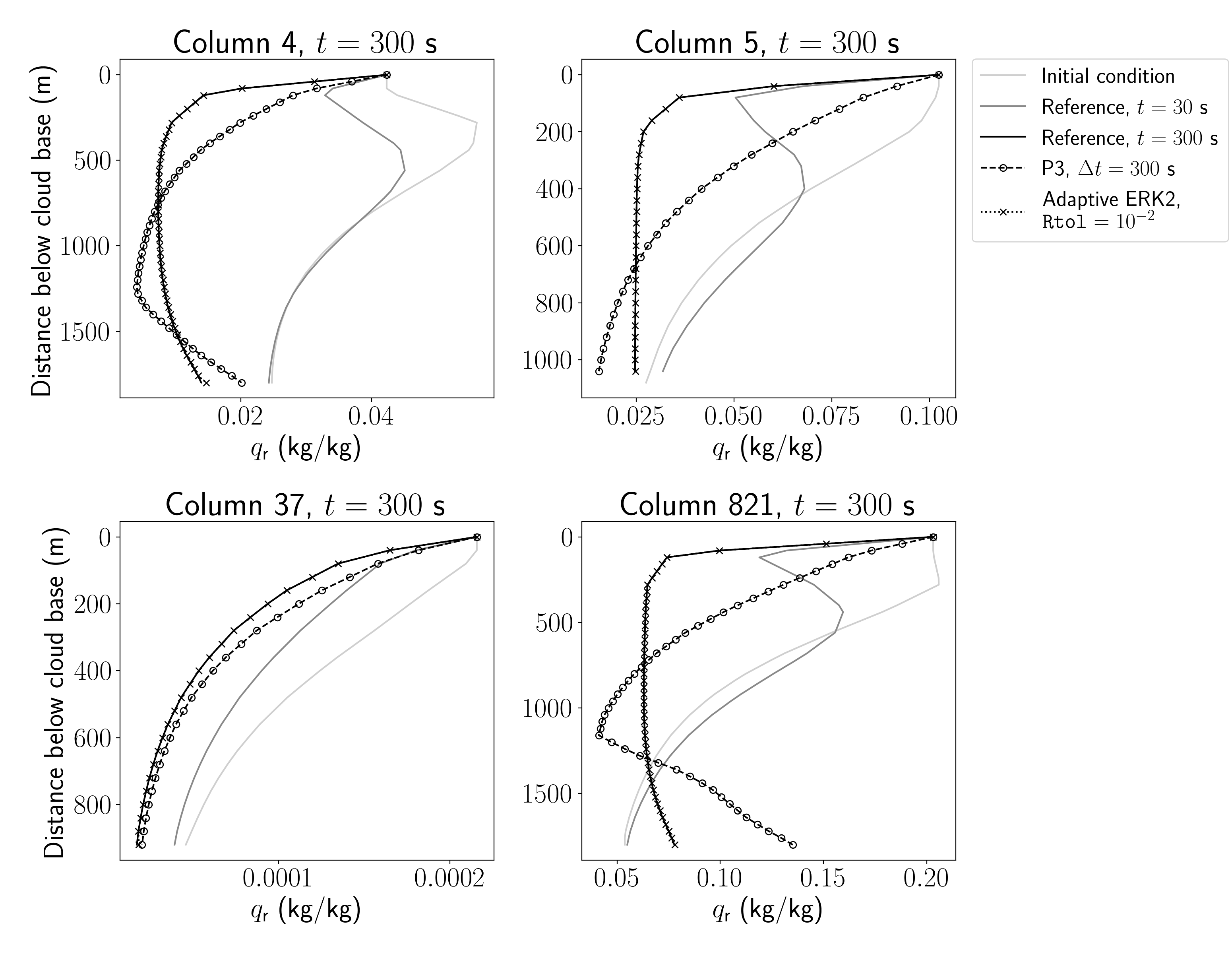}
    \caption{Selected solution profiles at $t=\SI{300}{\second}$ using the adaptive second-order ERK method with $\texttt{Rtol} = 10^{-2}$. Comparisons with the P3 method with $\Delta t = \SI{300}{\second}$, the initial condition from E3SM data, and the highly-resolved reference solution at $t = \SI{30}{\second}$ and $\SI{300}{\second}$ are also provided.}
    \label{fig:SUNDIALS profiles}
\end{figure}

To demonstrate the qualitative improvement in the numerical solutions over those produced by the P3 method, Figure \ref{fig:SUNDIALS histogram} plots a histogram of the pointwise relative differences for the second-order ERK method solutions with adaptive time step using the largest relative tolerance of \num{1.e-2} compared with the highly-resolved reference solution. When compared to the pointwise relative difference histogram for the P3 scheme in Figure \ref{fig:P3 histogram} we see a significant improvement in the error distribution. In turn, this can result in substantial changes in the solution. For instance, we see in Figure \ref{fig:SUNDIALS profiles} for a selection of columns that the second-order ERK method is visually indistinguishable from the high-resolution reference solution while P3 often does not correctly capture features such as boundary layers. 

MRI-GARK methods are less efficient than their explicit counterparts (i.e., take slightly more wall clock time to achieve a given solution accuracy) for our context, since the former perform best when there is a sufficiently large separation of process timescales. Recall that the partitioning of the processes for the MRI-GARK methods is based on the analysis in Figure \ref{fig:process rates} which shows that sedimentation usually has the fastest timescale. However, there are two key scenarios to note: (1) there are columns where evaporation and/or self-collection have faster timescales than sedimentation, and (2) in columns where the sedimentation timescale is fastest, it may only be less than an order of magnitude faster than those of the other processes in some columns while in others, it is more than three orders of magnitude faster. The wide spread of timescale differences means that overall, treating sedimentation as a fast process and evaporation and self-collection as slow processes is not more efficient than treating all processes with the same time step size (as in ERK methods). In future work with additional processes included, we expect there to be a greater separation of timescales which may yield improved results with MRI-GARK methods. As with the fixed step results in Section \ref{sec:fixed step results}, DIRK and ImEx methods are not competitive with the explicit methods when used in our rainshaft model.

Based on the results in this section, we recommend the second-order ERK method with adaptive time stepping as an alternative to the P3 scheme whenever a WRMSE of $\sim\SI{1.e-7}{\kilogram\per\kilogram}$ or greater is desired. For WRMSE less than $\SI{1.e-7}{\kilogram\per\kilogram}$, we recommend the third-order ERK method. In practice, we expect uncertainties from other sources to be far higher than this threshold (e.g. uncertainties in the tuning and structure of the P3 parameterization itself or due to coarse spatial resolution), so $\SI{1.e-7}{\kilogram\per\kilogram}$ is likely a far smaller tolerance for error than will be needed in most cases. However, this does serve as a reminder that the most efficient method does depend to an extent on one's tolerance for error in a simulation.

%
%
\section{Conclusion} \label{sec:conclusion}

We have constructed a rainshaft model by extracting the sedimentation, evaporation, and self-collection parameterizations from the P3 microphysics scheme, as implemented in E3SMv3. Analytically derived timescales for these processes show that all three processes are associated with fast ($<\SI{10}{\second}$) timescales for at least some conditions present in E3SM. For sedimentation, the shortest timescales are derived from the CFL condition, rather than physical, grid-independent timescales. The current E3SM microphysics time step (\SI{300}{\second}) is inadequate to resolve these processes accurately. The $\sim\SI{0.4}{\second}$ time step required for a reasonably accurate answer would greatly increase the computational cost of P3 (e.g. by a factor of \num{40} for the rainshaft model) relative to the default P3 scheme with \SI{300}{\second} time step.

Using the newly-developed SPAECIES framework, we have tested a wide variety of potential alternative time integration methods, using implementations in the SUNDIALS library to accomplish this. All of the tested second- and higher-order methods produce significantly lower estimated error in $q_{\text{r}}$ than P3's current first-order method for short ($\leqslant\SI{1}{\second}$) time steps, with ERK methods being the most efficient. By leveraging adaptive time stepping, which dynamically changes the time step to keep estimated time integration error below a given target, we are able to reduce the wall clock time of these methods by roughly a factor of \num{6} while achieving the same level of accuracy as the same methods with fixed time stepping. These methods do not rely on limiters or an explicitly calculated CFL condition based on a single process (i.e., sedimentation) for stability.

For the three processes under consideration in the rainshaft model, we recommend that the second and third-order ERK methods with adaptive time stepping be used in place of the current P3 scheme. Relative to the default P3 scheme with \SI{300}{\second} time step, these methods reduce the estimated time integration error by at least \num{100}x while increasing wall-clock time by only a factor of \num{2.6}x. This is \num{15}x faster than a simulation where P3 scheme's time step is reduced to \SI{0.4}{\second} to obtain that same level of accuracy. The other methods investigated (DIRK, ImEx, MRI-GARK) were found to be less efficient for this particular model, though the timescale analysis carried out in this work suggests that they may be more efficient for other models (e.g. with a different resolution, or with additional slower processes). In the case of MRI-GARK methods, there is no fixed partitioning of the microphysical processes which correctly captures the most significant timescale differences across all columns. For DIRK and ImEx methods, we find that sedimentation has only a modestly faster timescale for stability than for accuracy, which itself is only modestly faster than the timescales for the local microphysical processes (or in some cases slower). This timescale analysis suggests that sedimentation is only a mildly stiff process which still requires a moderately small time step ($<\SI{3}{\second}$) to return accurate solutions, and therefore DIRK and ImEx methods cannot improve efficiency simply by taking very large time steps. This analysis provides a blueprint for anticipating the performance of various time integration methods, which we intend to apply to additional microphysical processes in future work.
\begin{table}[ht!]
    \centering
    \begin{tabular}{p{3.5cm}|p{3.5cm}|p{3.5cm}}
        Metric & P3 scheme & Optimal SUNDIALS scheme\\ \hline
        Time to resolve important solution features & \SI{8.47}{\second} & \SI{0.563}{\second} (15x speedup) \\ \hline
        Fastest time to solution & \SI{0.215}{\second} & \SI{0.563}{\second} (2.6x slower)\\ \hline
        Relative WRMSE at fastest time to solution & \num{78.4}\% & \num{0.733}\% (107x reduction)
    \end{tabular}
    \caption{A summary of key performance metrics for the default P3 scheme and the optimal SUNDIALS scheme (second-order ERK).}
    \label{tab:results summary}
\end{table}

\subsection{Implications for Fully Coupled Models} \label{sec:implications}

We conclude with three points about the extent to which our rainshaft model study has implications for fully coupled models like E3SM. First, we are expanding our SPAECIES implementation to cover all P3 processes, and it is likely that the specific methods that perform best will be different when the full cloud and ice physics is implemented. E.g., the full set of processes may have a very different range of time scales, affecting the relative rankings of ERK, DIRK, ImEx, and MRI-GARK methods. Nevertheless, we hypothesize that adaptive time stepping will continue to outperform methods with a fixed time step; intuitively we would expect a model with a wider range of meteorological conditions to benefit even more from adapting the time step to those conditions.

Second, we believe an increase in computational cost on the order of $2.6\times$ for rain microphysics is, though significant, not unreasonably expensive in the context of a global model where microphysics is still a small fraction of overall cost. In EAMv3 at \SI{1}{\degree} effective resolution, the entire P3 microphysics requires about 10\% of the total run time \cite{E3SMv3Overview}, and in the $\sim$\SI{3}{\kilo\meter} variant of E3SM, all the physics parameterizations put together use about 15\% of total run time \cite{donahue2024scream}.

Third, although this study focuses on the error incurred when integrating the microphysics alone over a single \SI{300}{\second} time step (as is done in E3SMv3's substepping), previous studies have shown that time integration errors can noticeably affect precipitation rates in full atmosphere models. In \citeA{zheng2020magic}, reducing the microphysics time step from \SI{300}{\second} to \SI{30}{\second} in a single-column E3SMv1 case dramatically reduced the amount of rain mass present below cloud base, reducing rain evaporation and increasing precipitation efficiency. In \citeA{santos2021e3sm}, reducing the microphysics step size to \SI{10}{\second} increased the large-scale precipitation rate in E3SMv1 global runs by 10\%; this effect was also suspected (though not confirmed) to be mediated by rain evaporation rates. These studies suggest that changes to time integration can have effects comparable to changes in the parameterization formulation or tuning of microphysics schemes, and we believe this is important to consider for future model development.

\section*{Open Research Section}

Version 1.0.0 of SPAECIES is used to produce the output data in this manuscript \cite{SPAECIESv1}. SPAECIES is developed on GitHub at \url{https://github.com/pnnl/SPAECIES} and is released under a BSD-3 open source license. The E3SMv3 data used to formulate the boundary and initial conditions described in Section \ref{sec:boundary conditions} may be found on Zenodo \cite{inputdataset}. Output data from SPAECIES, which includes temperature, water vapor mixing ratio, raindrop number concentration, and rain mass mixing ratio profiles at \SI{300}{\second} using the time integration methods described in Section \ref{sec:time integration}, may also be found on Zenodo \cite{outputdataset} alongside the postprocessing scripts used to produce the figures in this manuscript. Both input and output data are made available under the Creative Commons Attribution 4.0 International license.

\acknowledgments
This work was performed under the auspices of the U.S. Department of Energy by Lawrence Livermore National Laboratory under Contract DE-AC52-07NA27344. LLNL-JRNL-2013883-DRAFT.
Pacific Northwest National Laboratory is operated for the U.S. Department of Energy by Battelle Memorial Institute under contract DE-AC06-76RLO1830.
This research was primarily supported by the U.S. Department of Energy, Office of Science, Office of Advanced Scientific Computing Research and Office of Biological and Environment Research (BER), Scientific Discovery through Advanced Computing (SciDAC) program and through
ASCR via the Frameworks, Algorithms, and Scalable Technologies for Mathematics (FASTMath) SciDAC Institute.

The authors declare there are no conflicts of interest for this manuscript.

%
%

\bibliography{main}

%
%
%
%
%

\appendix
\section{Microphysical Processes} \label{app:processes}

Note that in the original P3 implementation, all processes are disabled (rates and fall speeds set to zero) in grid cells where $q_{\text{r}}<q_{\text{small}}$. In our rainshaft model, regularization of the mean diameter (see Appendix \ref{app:lambdar regularization}) means that sedimentation is now always enabled. The rain evaporation tendency is also regularized (see Appendix \ref{app:evap regularization}) and is always active. Lastly, the raindrop self-collection (coalescence and breakup) tendency is always active in our model. 

\subsection{Raindrop self-collection}

The raindrop self-collection process $S_{\text{rsc}}$ is given by
\begin{equation} \label{eq:rsc tendency}
    S_{\text{rsc}}(T,q,n_{\text{r}},q_{\text{r}}) = \SI{5.78}{\meter\tothe{3}\per\kilo\gram\per\second} \cdot \min\{1, B(n_{\text{r}},q_{\text{r}})\} \cdot n_{\text{r}}q_{\text{r}}\rho(T,q),
\end{equation}
where $B$ is the breakup function defined by
\begin{align} \label{eq:breakup}
    B(n_{\text{r}},q_{\text{r}}) = 2 - \exp\left\{\SI{2300}{\meter\tothe{-1}} \left(\frac{1}{\lambda_{\text{r}}(n_{\text{r}}, q_{\text{r}})} - \SI{2.8e-4}{\meter}\right)\right\}.
\end{align}

\subsection{Rain evaporation}

The evaporation mass rate $S_{q,\text{evap}}$ is given by
\begin{align} \label{eq:evap tendency}
    S_{q,\text{evap}}(T,q,n_{\text{r}},q_{\text{r}}) = \begin{dcases}
        \frac{q_{\text{sl}}(T) - q}{a_{\text{bl}}(T)\tau(T,q,n_{\text{r}},q_{\text{r}})}, &\text{if}\;q_{\text{sl}}(T) > q\\
        0, &\text{else},
    \end{dcases}
\end{align}
in terms of the saturation water vapor mass mixing ratio $q_{\text{sl}}$, the psychrometric correction factor $a_{\text{bl}}$, and timescale $\tau$. These quantities and others necessary for their computation are defined in Table \ref{tab:evap equations}. The evaporation number rate is calculated assuming that evaporation does not change mean drop size and is given by
\begin{align}
    S_{n,\text{evap}}(T,q,n_{\text{r}},q_{\text{r}}) = \frac{n_{\text{r}}}{q_{\text{r}}} S_{q,\text{evap}}(T,q,n_{\text{r}},q_{\text{r}}).
    \label{eq:evaporation-number-rate}
\end{align}
\begin{table}[t!]
    \centering
    \let\displaystyle\textstyle
    \begin{tabular}{p{1.5cm} | p{4cm} | p{4cm}}
        Symbol & Quantity & Value\\
        \hline 
        $\epsilon_{\text{H2O}}$ & Water vapor/dry air molecular mass ratio & \num{0.62197}\\
        \hline 
        $L_{\text{v}}$ & Latent heat of vaporization for water & \SI{2.501e6}{\joule\per\kilo\gram}\\
        \hline 
        $c_{p}$ & Heat capacity of dry air at constant pressure & \SI{1.00464e3}{\joule\per\kelvin\per\kilo\gram}\\
        \hline 
        $\rho_{\text{w}}$ & Density of water & \SI{1000}{\kilo\gram\per\meter\tothe{3}} \\
        \hline 
        $R_{\text{d}}$ & Gas constant for dry air & \SI{287.04}{\joule\per\kelvin\per\kilo\gram}\\
        \hline 
        $R_{\text{v}}$ & Gas constant for water vapor & \SI{461.50}{\joule\per\kelvin\per\kilo\gram}
    \end{tabular}
    \caption{Physical constants and their values used in the rainshaft model.}
    \label{tab:constants}
\end{table}

\begin{table}[t!]
    \centering
    \let\displaystyle\textstyle
    \begin{tabular}{p{3cm} | p{10cm}}
        Quantity & Definition  \\
        \hline
        Saturation water vapor mass mixing ratio & \( q_{\text{sl}}(T) = \frac{\epsilon_{\text{H2O}} e_{sl}(T)}{p - e_{sl}(T)} \)\\
        \hline
        Saturation vapor pressure & \begin{minipage}{0.25\textwidth}
            Given implicitly by
            \begin{align*}
                \log{e_{sl}(T)} = 54.842763 - \frac{6763.22}{T} - 4.21 \log\left(T\right) + 0.000367 T + \\
    \left[53.878 - \frac{1331.22}{T} - 9.44523 \log\left(T\right) + 0.014025 T\right]\cdot\\
    \tanh\left[0.0415 \left(T - 218.8\right)\right]
            \end{align*}
        \end{minipage}\\
        \hline 
        Psychrometric correction factor & \( a_{\text{bl}}(T) = 1 + \frac{L_\text{v}^2 q_{\text{sl}}(T)}{c_p R_{\text{v}} T^2} \)\\
        \hline
        Timescale & \begin{minipage}{0.25\textwidth}
            \(\begin{aligned}
                \tau(T,q,n_{\text{r}},q_{\text{r}}) = \left[2 \pi n_{\text{r}} \rho(T,q) D_v(T) \left(\frac{0.78}{\lambda_{\text{r}}(n_{\text{r}},q_{\text{r}})} + \right.\right.\\
                \left.\left.0.32 S_c^{1/3}(T,q) \sqrt{\frac{\rho(T,q)}{\mu_{\text{visc}}(T)}} V_{\text{evap}}(\lambda_{\text{r}}(n_{\text{r}},q_{\text{r}}))\right)\right]^{-1} 
            \end{aligned} \)
        \end{minipage}\\
        \hline 
        Water vapor diffusivity & \( D_{v}(T) = \SI{8.794e-5}{\meter\tothe{2}\pascal\per\second\per\kelvin\tothe{1.81}} \cdot \frac{T^{1.81}}{p} \)\\
        \hline 
        Dynamic viscosity of air & \( \mu_{\text{visc}}(T) = \qty[per-mode=fraction]{1.496e-6}{\kilo\gram\per\meter\per\second\per\kelvin\tothe{1/2}} \times \frac{T^{3/2}}{T + \SI{120}{K}} \)\\
        \hline 
        Schmidt number & \( S_{c}(T,q) = \frac{\mu_{\text{visc}}(T)}{\rho(T,q) D_v(T)} \)\\
        \hline 
        Scale parameter & \( \lambda_{\text{r}}(n_{\text{r}},q_{\text{r}}) = \left( \frac{\pi\rho_{\text{w}}n_{\text{r}}}{q_{\text{r}}} \right)^{1/3} \)\\
        \hline 
        Evaporation fall speed & \begin{minipage}{0.5\textwidth} 
        \(
            V_{\text{evap}}(\lambda_{\text{r}}) = \int_{0}^{\infty} \sqrt{D\cdot v(D)} \lambda_{\text{r}}De^{-\lambda_{\text{r}}D}\;dD,
        \)
        
        where $v(D)$ is defined as
        \begin{align*}
            v(D) = \begin{cases}
                \num{4579.5} \times \left[m_{\text{cgs}}(D)\right]^{2/3} \;\text{if } D \leqslant \SI{134.43}{\micro\meter} \\
                \num{49.62} \times \left[m_{\text{cgs}}(D)\right]^{1/3}  \;\text{if } \SI{134.43}{\micro\meter} < D \leqslant \SI{1511.64}{\micro\meter} \\
                \num{17.32} \times \left[m_{\text{cgs}}(D)\right]^{1/6}  \;\text{if } \SI{1511.64}{\micro\meter} < D \leqslant \SI{3477.84}{\micro\meter} \\
                \num{9.17}  \;\text{if } \SI{3477.84}{\micro\meter} < D
            \end{cases}
        \end{align*}
        and \( m_{\text{cgs}}(D) = \SI{1000}{\gram\per\kilogram} \times \frac{\pi \rho_{\text{w}}}{6} D^3 \)
        \end{minipage}\\
    \end{tabular}
    \caption{Quantities required for computation of the mass and number rates.}
    \label{tab:evap equations}
\end{table}

\subsection{Sedimentation fall speeds}

The sedimentation fall speeds $v_{n_{\text{r}}}$ and $v_{q_{\text{r}}}$ are defined in terms of the scale parameter $\lambda_{\text{r}}$ as
\begin{equation}
    \begin{aligned}
        v_{n_{\text{r}}}(T,q,n_{\text{r}},q_{\text{r}}) &= \left(\frac{\SI{1.e5}{\pascal}}{\rho(T,q) R_{\text{d}} \times \SI{273.15}{\kelvin}}\right)^{0.54} \frac{\int_{0}^{\infty} v(D)e^{-\lambda_{\text{r}}(n_{\text{r}},q_{\text{r}})D}\;dD}{\int_{0}^{\infty} e^{-\lambda_{\text{r}}(n_{\text{r}},q_{\text{r}})D}\;dD}\\
        v_{q_{\text{r}}}(T,q,n_{\text{r}},q_{\text{r}}) &= \left(\frac{\SI{1.e5}{\pascal}}{\rho(T,q) R_{\text{d}} \times \SI{273.15}{\kelvin}}\right)^{0.54} \frac{\int_{0}^{\infty} v(D)D^{3}e^{-\lambda_{\text{r}}(n_{\text{r}},q_{\text{r}})D}\;dD}{\int_{0}^{\infty} D^{3}e^{-\lambda_{\text{r}}(n_{\text{r}},q_{\text{r}})D}\;dD},
    \end{aligned}
\end{equation}
where $v(D)$ and $\lambda_{\text{r}}(n_{\text{r}},q_{\text{r}})$ are defined as in Table \ref{tab:evap equations}.

\section{Derivation of Hyperbolic Characteristic Speeds for the Rainshaft Model} \label{app:hyperbolics}

Recall that the value of $v_{q_{\text{r}}}$ has traditionally been used for the characteristic sedimentation speed $\tilde{v}$, as it is the faster of the two fall speeds $v_{n_{\text{r}}}$ and $v_{q_{\text{r}}}$.
The better choice for the characteristic sedimentation speed is that from the characteristic curves from analyzing the hyperbolic conservation law associated with sedimentation.
The rainshaft model (Equation \ref{eq:rainshaft model}) is not expressed in conservation law form, and expressing it as such requires deriving it from a set of conservation laws.
Consider the 1D system of equations for the conservation of water vapor mass, rain droplet number, and rain droplet mass in air:
\begin{equation}
  \begin{dcases}
      \frac{\partial}{\partial t} (\rho q) + \frac{\partial}{\partial z} (\rho q w) = \rho S_{q,\text{evap}} \\
      \frac{\partial}{\partial t} (\rho n_{\text{r}}) + \frac{\partial}{\partial z} (\rho n_{\text{r}} w - \rho n_{\text{r}} v_{n_{\text{r}}} ) = -\rho S_{n,\text{evap}} - \rho S_{\text{rsc}} \\
      \frac{\partial}{\partial t} (\rho q_{\text{r}}) + \frac{\partial}{\partial z} (\rho q_{\text{r}} w - \rho q_{\text{r}} v_{q_{\text{r}}} ) = - \rho S_{q,\text{evap}}.
  \end{dcases}
  \label{eq:conservation_water}
\end{equation}
Here, $\rho$ denotes the air mass, $w$ denotes the air speed, and $q$, $n_\text{r}$, $q_\text{r}$, $v_{n_\text{r}}$, $v_{q_\text{r}}$, $S_{q,\text{evap}}$, $S_{n,\text{evap}}$ and $S_\text{rsc}$ all retain their meaning from the rainshaft model. Note that Equation \ref{eq:conservation_water} is the subset of the full 3D system that excludes the horizontal motion terms.  We focus on the 1D system because those horizontal motion terms are typically handled by a dynamical core or other parameterizations separate from the microphysics scheme.

The air mass $\rho$, speed $w$, and temperature $T$ are governed by the following system of equations for the conservation of air mass, momentum, and energy:
\begin{equation}
  \begin{dcases}
    \frac{\partial}{\partial t}(\rho) + \frac{\partial}{\partial z} (\rho w) = 0 \\
    \frac{\partial}{\partial t}(\rho w) + \frac{\partial}{\partial z} (\rho w^2 + p ) = -\rho g  \\
    \frac{\partial}{\partial t} (E) + \frac{\partial}{\partial z} \big((E + p) w\big) = - \rho  c_v \frac{L_\text{v}}{c_p} S_{q,\text{evap}}
  \end{dcases}
  \label{eq:conservation_air}
\end{equation}
where $p$ is the pressure, $g$ is the gravitational constant, $E$ is the energy, and $c_v$ is the specific heat capacity at constant volume.
We assume the air follows the moist ideal gas law
\begin{equation}
  p = \rho R_{\text{d}} T_v(T,q) \text{, where } T_v(T,q) = \left( 1 + \frac{q}{\epsilon_{\text{H2O}}}\right)T,
  \label{eq:equation_of_state}
\end{equation}
where $R_{\text{d}}$ is the ideal gas constant, which closes the system in Equation \ref{eq:conservation_air} by determining the specific internal energy as $e = c_v T$ and therefore the energy $E = \rho c_v T + \tfrac{1}{2} \rho w^2$.
With this definition of $E$, along with the equations for the conservation of air mass and momentum, the conservation of energy equation can be expressed as
\begin{equation}
  \frac{\partial}{\partial t}(\rho T) + \frac{\partial}{\partial z}(\rho T w) = - \rho \frac{L_\text{v}}{c_p} S_{q,\text{evap}} -\frac {1}{c_v}\left(p \frac{\partial w}{\partial z} - \rho w g \right).
  \label{eq:temperature_equation}
\end{equation}

\subsection{Deriving the rainshaft model from the system of conservation laws}

  Given the system of conservation laws in Equation \ref{eq:conservation_water} and the temperature equation in Equation \ref{eq:temperature_equation}, we now derive the rainshaft model in Equation \ref{eq:rainshaft model}.
  First, the conservation of air mass equation in Equation \ref{eq:conservation_air} is leveraged to express the system in Equation \ref{eq:conservation_water} in the following form:
  \begin{equation}
    \begin{dcases}
      \frac{\partial q}{\partial t} + w \frac{\partial q}{\partial z} = S_{q,\text{evap}} \\
      \frac{\partial n_\text{r}}{\partial t} + w \frac{\partial n_\text{r}}{\partial z} - \frac{1}{\rho} \frac{\partial}{\partial z} (\rho n_{\text{r}} v_{n_{\text{r}}} ) = -S_{n,\text{evap}} - S_{\text{rsc}} \\
      \frac{\partial q_\text{r}}{\partial t} + w \frac{\partial q_\text{r}}{\partial z} - \frac{1}{\rho} \frac{\partial}{\partial z} (\rho q_{\text{r}} v_{q_{\text{r}}} ) = - S_{q,\text{evap}}
    \end{dcases}
  \end{equation}
  Note that if the $w  \tfrac{\partial}{\partial z}$ advection terms are operator split to be handled outside the cloud microphysics (e.g., by the dynamical core), then one arrives at the rainshaft model equations for $q$, $n_\text{r}$, and $q_\text{r}$.
  Similarly, the conservation of air mass equation in Equation \ref{eq:conservation_air} is leveraged to express the temperature equation (Equation \ref{eq:temperature_equation}) in the following form:
  \begin{equation}
    \frac{\partial T}{\partial t} + w \frac{\partial T}{\partial z} = - \frac{L_\text{v}}{c_p} S_{q,\text{evap}} - \frac{1}{c_v} \left(\frac{p}{\rho} \frac{\partial w}{\partial z} - w g \right).
  \end{equation}
  As with Equation \ref{eq:conservation_water}, note that the rainshaft model equation for $T$ is attained if the $w \tfrac{\partial T}{\partial z}$ (advection of temperature) and $\tfrac{p}{\rho} \tfrac{\partial w}{\partial z} - w g$ (residual of work done by vertical motion and change in potential energy) terms are operator split to be handled outside the cloud microphysics.
  Given the rainshaft model is derived from Equation \ref{eq:conservation_water} and Equation \ref{eq:temperature_equation} with some terms ignored by operator splitting, we can consider the remaining terms in Equation \ref{eq:conservation_water} and Equation \ref{eq:temperature_equation} as the conservation form of the rainshaft model (Equation \ref{eq:rainshaft model}),
  \begin{equation}
    \begin{dcases}
      \frac{\partial}{\partial t} \big(\rho(T,q) T\big) = - \rho \frac{L_{v}}{c_p} S_{q,\text{evap}} \\
      \frac{\partial}{\partial t} \big( \rho(T,q) q \big) = \rho S_{q,\text{evap}} \\
      \frac{\partial}{\partial t} \big( \rho(T,q) n_{\text{r}} \big) - \frac{\partial}{\partial z} \big(v_{n_{\text{r}}}(T,q,n_{\text{r}},q_{\text{r}}) \rho(T,q) n_{\text{r}} \big) = -\rho S_{n,\text{evap}} -\rho S_\text{rsc}  \\
      \frac{\partial}{\partial t} \big( \rho(T,q) q_{\text{r}} \big) - \frac{\partial}{\partial z} \big(v_{q_{\text{r}}}(T,q,n_{\text{r}},q_{\text{r}}) \rho(T,q) q_{\text{r}} \big) = - \rho S_{q,\text{evap}}
    \end{dcases}
    \label{eq:conservation_rainshaft}
  \end{equation}
where $\rho(T,q) = \frac{p}{R_{\text{d}} T_v(T,q)}$ from the moist ideal gas law (Equation \ref{eq:equation_of_state}).

\subsection{Deriving the hyperbolic characteristic speeds}
  To derive the hyperbolic characteristic speeds in Equation \ref{eq:conservation_rainshaft}, we first note the fall speeds are functions of $T$, $q$, $n_{\text{r}}$, and $q_{\text{r}}$, through $\rho(T,q)$ and $\lambda_{\text{r}}(n_{\text{r}},q_{\text{r}}) \equiv \lambda_{\text{r}}(\tfrac{n_{\text{r}}}{q_{\text{r}}})$:
  \begin{equation}
    v_{n_{\text{r}}}(T,q,n_{\text{r}},q_{\text{r}}) \equiv v_{n_{\text{r}}}\big(\rho(T,q),\lambda_{\text{r}} \big(\tfrac{n_{\text{r}}}{q_{\text{r}}} \big)\big)
      \text{ and }
      v_{q_{\text{r}}}(T,q,n_{\text{r}},q_{\text{r}}) \equiv v_{q_{\text{r}}}\big(\rho(T,q),\lambda_{\text{r}} \big( \tfrac{n_{\text{r}}}{q_{\text{r}}} \big)\big).
  \end{equation}
  Introduce $\hat{T} = \rho T$ and $\hat{q} = \rho q$, and note that the density $\rho(T,q)$ as defined in Equation \ref{eq:equation_of_state} can be written as $\rho(\hat{T},\hat{q})$:
  \begin{equation}
    p = \rho R_{\text{d}} T \left(1 + \frac{q}{\epsilon_{\text{H2O}}} \right) 
      \quad \Rightarrow \quad p = R_{\text{d}} \hat{T} \left(1 + \frac{\hat{q}}{\epsilon_{\text{H2O}}} \frac{1}{\rho}\right) 
      \quad \Rightarrow \quad \rho = \frac{\hat{q}/\epsilon_{\text{H2O}}}{\frac{p}{R_{\text{d}} \hat{T}} - 1}.
  \end{equation}
  Introduce $\hat{n}_{\text{r}} = \rho n_{\text{r}}$ and $\hat{q}_{\text{r}} = \rho q_{\text{r}}$ to fully express Equation \ref{eq:conservation_rainshaft} in terms of conserved quantities:
  \begin{equation}
    \begin{dcases}
      \frac{\partial \hat{T}}{\partial t} = - \rho \frac{L_{v}}{c_p} S_{q,\text{evap}} \\
      \frac{\partial \hat{q}}{\partial t} = \rho S_{q,\text{evap}} \\
      \frac{\partial \hat{n}_{\text{r}}}{\partial t} - \frac{\partial}{\partial z} \left(v_{n_{\text{r}}}\big(\rho(\hat{T},\hat{q}), \lambda_{\text{r}} \big(\tfrac{\hat{n}_{\text{r}}}{\hat{q}_{\text{r}}}\big) \big) \hat{n}_{\text{r}} \right) = -\rho S_{n,\text{evap}} -\rho S_\text{rsc}  \\
      \frac{\partial \hat{q}_{\text{r}}}{\partial t} - \frac{\partial}{\partial z} \left(v_{q_{\text{r}}}\big(\rho(\hat{T},\hat{q}), \lambda_{\text{r}} \big(\tfrac{\hat{n}_{\text{r}}}{\hat{q}_{\text{r}}}\big) \big) \hat{q}_{\text{r}} \right) = - \rho S_{q,\text{evap}}
    \end{dcases}
  \end{equation}
  Note the flux function is
  \begin{equation}
    \mathbf{f}(\hat{T},\hat{q},\hat{n}_{\text{r}},\hat{q}_{\text{r}}) = \left[0, 0, v_{n_{\text{r}}}\big(\rho(\hat{T},\hat{q}), \lambda_{\text{r}} \big(\tfrac{\hat{n}_{\text{r}}}{\hat{q}_{\text{r}}}\big)\big) \hat{n}_{\text{r}}, v_{q_{\text{r}}} \big(\rho(\hat{T},\hat{q}), \lambda_{\text{r}} \big(\tfrac{\hat{n}_{\text{r}}}{\hat{q}_{\text{r}}}\big)\big) \hat{q}_{\text{r}} \right],
\end{equation}
with the following Jacobian:
\begin{equation}
  \mathbf{J_f}(\hat{T},\hat{q},\hat{n}_{\text{r}},\hat{q}_{\text{r}}) = 
  \begin{bmatrix}
    0 & 0 & 0 & 0 \\
    0 & 0 & 0 & 0 \\
    \hat{n}_{\text{r}} \frac{\partial v_{n_{\text{r}}}}{\partial \rho} \frac{\partial \rho}{\partial \hat{T}} 
      & \hat{n}_{\text{r}} \frac{\partial v_{n_{\text{r}}}}{\partial \rho} \frac{\partial \rho}{\partial \hat{q}} 
        &  v_{n_{\text{r}}} + \frac{\partial v_{n_{\text{r}}}}{\partial \lambda_{\text{r}}} \lambda_{\text{r}}' \frac{\hat{n}_{\text{r}}}{\hat{q}_{\text{r}}}
          & - \frac{\partial v_{n_{\text{r}}}}{\partial \lambda_{\text{r}}} \lambda_{\text{r}}' \left( \frac{\hat{n_{\text{r}}}}{\hat{q}_{\text{r}}}\right)^2 \\
    \hat{q}_{\text{r}} \frac{\partial v_{q_{\text{r}}}}{\partial \rho} \frac{\partial \rho}{\partial \hat{T}} 
      & \hat{q}_{\text{r}} \frac{\partial v_{q_{\text{r}}}}{\partial \rho} \frac{\partial \rho}{\partial \hat{q}} 
        & \frac{\partial v_{q_{\text{r}}}}{\partial \lambda_{\text{r}}} \lambda_{\text{r}}'
          & v_{q_{\text{r}}} - \frac{\partial v_{q_{\text{r}}}}{\partial \lambda_{\text{r}}} \lambda_{\text{r}}' \frac{\hat{n}_{\text{r}}}{\hat{q}_{\text{r}}}
  \end{bmatrix}.
\end{equation}
The nonzero eigenvalues of the Jacobian are the eigenvalues of 
\begin{equation}
  \begin{bmatrix}
    v_{n_{\text{r}}} + \frac{\partial v_{n_{\text{r}}}}{\partial \lambda_{\text{r}}} \lambda_{\text{r}}' \frac{\hat{n}_{\text{r}}}{\hat{q}_{\text{r}}}
      & - \frac{\partial v_{n_{\text{r}}}}{\partial \lambda_{\text{r}}} \lambda_{\text{r}}' \left( \frac{\hat{n_{\text{r}}}}{\hat{q}_{\text{r}}}\right)^2 \\
    \frac{\partial v_{q_{\text{r}}}}{\partial \lambda_{\text{r}}} \lambda_{\text{r}}'
      & v_{q_{\text{r}}} - \frac{\partial v_{q_{\text{r}}}}{\partial \lambda_{\text{r}}} \lambda_{\text{r}}' \frac{\hat{n}_{\text{r}}}{\hat{q}_{\text{r}}}
  \end{bmatrix}.
\end{equation}
Noting that $\lambda_{\text{r}}' \frac{\hat{n}_{\text{r}}}{\hat{q}_{\text{r}}} = \frac{1}{3}\lambda_{\text{r}}$, the characteristic polynomial for the eigenvalues $\sigma_{\text{sed}}$ is
\begin{equation}
  \left(v_{n_{\text{r}}} + \frac{\partial v_{n_{\text{r}}}}{\partial \lambda_{\text{r}}} \frac{\lambda_{\text{r}}}{3} - \sigma\right)
    \left(v_{q_{\text{r}}} - \frac{\partial v_{q_{\text{r}}}}{\partial \lambda_{\text{r}}} \frac{\lambda_{\text{r}}}{3} - \sigma\right) + \frac{\partial v_{n_{\text{r}}}}{\partial \lambda_{\text{r}}}\frac{\partial v_{q_{\text{r}}}}{\partial \lambda_{\text{r}}} \left(\frac{\lambda_{\text{r}}}{3}\right)^2 = 0.
    \label{eq:characteristic polynomial}
\end{equation}
The eigenvalues $\sigma_{\text{sed}}$, which are the roots of Equation \ref{eq:characteristic polynomial}, are the hyperbolic characteristic speeds in Equation \ref{eq:conservation_rainshaft}.

\section{Limiters in P3} \label{app:limiters}

\subsection{Positivity limiters}

Given a solution state $(T,q,n_{\text{r}},q_{\text{r}})$, the positivity limiter may be described by the mapping $\mathfrak{L}_{\text{pos}}$ defined by
\begin{align} \label{eq:positivity limiter}
    \mathfrak{L}_{\text{pos}}(T,q,n_{\text{r}},q_{\text{r}}) = \begin{dcases}
        \Big( \max\left\{T-q_{\text{r}}\frac{L_{v}}{c_{p}},0\right\}, \max\{q+q_{\text{r}},0\}, n_{\text{r}}, 0 \Big), &\text{if}\;q_{\text{r}} < q_{\text{small}}\\
        \Big( \max\{T,0\}, \max\{q,0\}, n_{\text{r}}, q_{\text{r}} \Big), &\text{else}.
    \end{dcases}
\end{align}
This limiter is designed to ensure that if $q_{\text{r}}$ becomes negative due to evaporation, then its value is reset to \num{0}, and the difference is taken from $q$ to conserve total water mass. Additionally, a simpler positivity is applied after sedimentation, defined by
\begin{align} \label{eq:sedimentation positivity limiter}
    \mathfrak{L}_{\text{sedpos}}(T,q,n_{\text{r}},q_{\text{r}}) =
        \Big( \max\{T,0\}, \max\{q,0\}, \max\{n_{\text{r}},0\}, \max\{q_{\text{r}},0\} \Big)
\end{align}
Since substepping in P3 should prevent the creation of substantially negative values, this limiter exists solely to handle very small negative values resulting from the inexact nature of floating point operations.

\subsection{Size limiter}

The size limiter ensures that mean drop size is kept between $D_{\text{min}}(= \SI{10}{\micro\meter})$ and $D_{\text{max}}(= \SI{500}{\micro\meter})$. This is accomplished by computing the corresponding rain numbers for $D_{\text{min}}$ and $D_{\text{max}}$:
\begin{align}
    n_{\text{r,min}} := \mathcal{F}(D_{\text{max}}) \cdot q_{\text{r}}, \quad n_{\text{r,max}} := \mathcal{F}(D_{\text{min}}) \cdot q_{\text{r}}.
\end{align}
Here, the conversion function $\mathcal{F}$ is defined by
\begin{align}
    \mathcal{F}(D) = \frac{6\left( \frac{\mu+1}{D} \right)^{3}}{\pi\rho_{\text{w}}(\mu+1)(\mu+2)(\mu+3)},
\end{align}
where $\mu=0$ is the shape parameter of the underlying gamma distribution for the rain number. The size limiter may then be described by the mapping $\mathfrak{L}_{\text{size}}$ defined by
\begin{align} \label{eq:size limiter}
    \mathfrak{L}_{\text{size}}(T,q,n_{\text{r}},q_{\text{r}}) = \Big( T,q, \max\{ \min\{ n_{\text{r}}, n_{\text{r,max}} \}, n_{\text{r,min}} \}, q_{\text{r}} \Big).
\end{align}
Given a vector of discrete solution states $\bm{y}$ defined in each grid cell of a column, we use the notation $\mathfrak{L}_{\text{pos}}(\bm{y})$ and $\mathfrak{L}_{\text{size}}(\bm{y})$ to indicate that the limiters are applied to the solution states in each grid cell.

\section{Regularization} \label{app:regularization}

\subsection{Mean diameter} \label{app:lambdar regularization}

At present, a limiter is applied to $\lambda_{\text{r}}$ to prevent the case when $q_{\text{r}} \to 0$, which would otherwise result in singular behavior of $\lambda_{\text{r}}$. However, limiters in P3 introduce a jump discontinuity with respect to $q_{\text{r}}$ as seen in the following definition:
\begin{equation}
    \lambda_{\text{r},\text{limited}}(n_{\text{r}},q_{\text{r}}) = \begin{dcases}
        \left( \frac{\pi\rho_{\text{w}}n_{\text{r}}}{q_{\text{r}}} \right)^{1/3}, &q_{\text{r}} > q_{\text{small}}\\
        0, &\text{else}.
    \end{dcases}
\end{equation}

There are several ways to address the jump discontinuity in $\lambda_{\text{r}}$ which prevent blow up while also maintaining continuity of $\lambda_{\text{r}}$ with respect to $q_{\text{r}}$. The simplest approach is to add a small tolerance to the denominator of $\lambda_{\text{r}}$ to prevent it from taking on zero values. We select the tolerance based on the value of $q_{\text{small}}$:
\begin{equation}
    \tilde{\lambda}_{\text{r}}(n_{\text{r}}, q_{\text{r}}) := \left( \frac{\pi\rho_{\text{w}}(n_{\text{r}} + q_{\text{small}}\cdot 10^{8})}{q_{\text{r}} + q_{\text{small}}} \right)^{1/3}.
\end{equation}
We select the value $q_{\text{small}} = 10^{-18}$ for the numerical experiments in this work. The impact of the regularization of $\lambda_{\text{r}}$ is evaluated alongside the impact of the evaporation tendency regularization in the next section.

\subsection{Evaporation tendency} \label{app:evap regularization}

It is straightforward to verify that despite its piecewise definition, the evaporation tendency in Equation \ref{eq:evap tendency} is continuous at $q = q_{\text{sl}}(T)$. However, its derivative contains a jump discontinuity at the same location. This discontinuity impacts the convergence of Jacobian-based nonlinear solvers such as Newton's method in DIRK methods which require continuity of the tendency Jacobians. 

We elect to introduce a seventh degree polynomial interpolation of $S_{q,\text{evap}}$ in the interval $q_{\text{sl}}(T) - \varepsilon \leqslant q \leqslant q_{\text{sl}}(T)$ with $0 < \varepsilon < q_{\text{sl}}(T)$ in order to enforce $C^{3}$ continuity of the evaporation tendency. This smoothness is required for the third-order time integration methods explored in Section \ref{sec:time integration}. In particular, we define
\begin{equation}
    \tilde{S}_{q,\text{evap}}(T,q,n_{\text{r}},q_{\text{r}}) = S_{q,\text{evap}}(T,q,n_{\text{r}},q_{\text{r}}) \cdot \begin{dcases}
        1, &q < q_{\text{sl}}(T) - \varepsilon\\
        Q(T,q,n_{\text{r}},q_{\text{r}}), &q_{\text{sl}}(T) - \varepsilon \leqslant q < q_{\text{sl}}(T)\\
        0, &q \geqslant q_{\text{sl}}(T),
    \end{dcases}
\end{equation}
where $Q$ is the seventh degree polynomial 
\begin{align}
    \begin{aligned}
        Q(T,q,n_{\text{r}},q_{\text{r}}) = &35\left( \frac{q-q_{\text{sl}}(T)}{\varepsilon} \right)^{4} + 84\left( \frac{q-q_{\text{sl}}(T)}{\varepsilon} \right)^{5} + 70\left( \frac{q-q_{\text{sl}}(T)}{\varepsilon} \right)^{6}\\
        &+ 20\left( \frac{q-q_{\text{sl}}(T)}{\varepsilon} \right)^{7}.
    \end{aligned}
\end{align}
%
%
We select the value of the regularization parameter $\varepsilon$ as $\varepsilon = (10^{-10})q_{\text{sl}}(T)$ for all simulations in this paper. 

\begin{figure}
    \centering
    \includegraphics[width=0.85\linewidth]{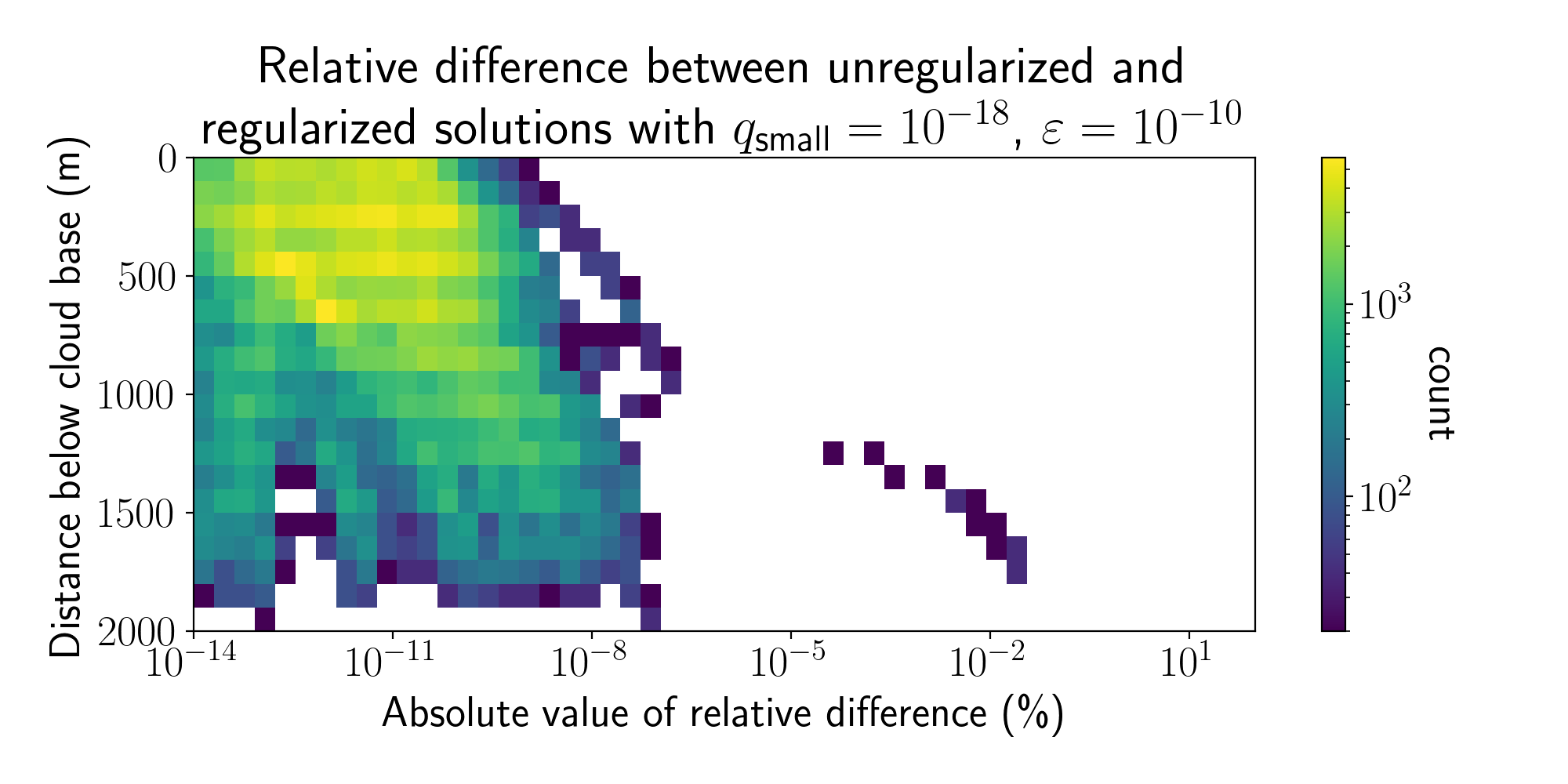}
    \caption{2D histogram of percent relative difference between highly-resolved solutions with and without regularization.}
    \label{fig:regularization error}
\end{figure}

To demonstrate that the regularizations utilized in this work do not substantially impact the solution behavior, a histogram of the pointwise relative difference between highly-resolved solutions obtained from the rainshaft model without and with regularization of $\lambda_{\text{r}}$ and $S_{q,\text{evap}}$ is provided in Figure \ref{fig:regularization error}. A fifth-order ERK method with $\Delta t = \num{1.e-2}$ is used to obtain these highly-resolved solutions. Points in all columns of the regularized solution have a relative difference error of less than \num{1.e-7}\% except for column 193 in the dataset, which has relative difference between \num{1.e-5} and \num{1.e-2}\%; we speculate that the larger difference in this column is due to the fact that the limiter in Equation \ref{eq:evap tendency} is particularly active here, thus resulting in solution values that more frequently fall in the cubic interpolation region of the regularized tendency.

\section{Time Integration Methods} \label{app:time integration methods}

\subsection{Table of Methods}

\begin{table}[ht!]
    \centering
    \scriptsize
    \begin{tabular}{c|ccc}
        Method Type & Order 1 & Order 2 & Order 3 \\ \hline
        Operator & Lie--Trotter & Strang \cite{strang1968construction} & Suzuki \cite{suzuki1992general} \\
        Splitting \\ \hline
        Explicit & --- & Based on Ralston's 2\textsuperscript{nd} Order Method & Bogacki--Shampine \\
        Runge--Kutta & & \cite{ralston1962runge} & \cite{bogacki1989pair} \\ \hline
        Diagonally Implicit & --- & Implicit part of ARK2 & ESDIRK3(2)5L[2]SA \\
        Runge--Kutta & & \cite{giraldo2013implicit} & \cite{kennedy2016diagonally} \\ \hline
        Implicit-Explicit & --- & ARK2 & ARK3(2)4L[2]SA \\
        Runge--Kutta & & \cite{giraldo2013implicit} & \cite{kennedy2003additive} \\ \hline
        MRI-GARK & --- & MRI-GARK-ERK22b & MRI-GARK-ERK33a \\
        & & \cite{sandu2019class} & \cite{sandu2019class} \\
    \end{tabular}
    \caption{Names and citations of integrators used in the numerical experiments. These correspond to the default integrators for the given orders in SUNDIALS 7.3.0.}
    \label{tab:methods}
\end{table}

\end{document}